\documentclass[%
 aip,
 amsmath,amssymb,
 reprint,%
]{revtex4-1}

\usepackage{graphicx}
\usepackage{dcolumn}
\usepackage{bm}
\usepackage[utf8]{inputenc}
\usepackage[hidelinks,breaklinks,colorlinks=true,linkcolor=blue,citecolor=blue,urlcolor=blue]{hyperref}
\usepackage[T1]{fontenc}
\usepackage{mathptmx}

\usepackage{etoolbox}
\makeatletter
\let\old@makecaption=\@makecaption
\usepackage{subcaption}
\let\@makecaption=\old@makecaption
\makeatother
\makeatletter
\def\@email#1#2{%
 \endgroup
 \patchcmd{\titleblock@produce}
  {\frontmatter@RRAPformat}
  {\frontmatter@RRAPformat{\produce@RRAP{*#1\href{mailto:#2}{#2}}}\frontmatter@RRAPformat}
  {}{}
}%
\makeatother
\begin{document}

\preprint{AIP/123-QED}

\author{Nikhil R. Agrawal}
\affiliation{ 
Department of Chemical and Biomolecular Engineering, University of California, Berkeley, California 94720, USA
}%

\author{Carlo Carraro}
\affiliation{ 
Department of Chemical and Biomolecular Engineering, University of California, Berkeley, California 94720, USA
}
\author{Rui Wang}
\email{ruiwang325@berkeley.edu}
\affiliation{ 
Department of Chemical and Biomolecular Engineering, University of California, Berkeley, California 94720, USA
}
\affiliation{ 
Materials Sciences Division, Lawrence Berkeley National Laboratory, Berkeley, California 94720, USA
}

\title{Understanding long-range opposite charge repulsion in multivalent salt solutions}

\date{\today}

\begin{abstract}
The electrostatic correlations between ions profoundly influence the structure and forces within electrical double layers. Here, we apply the modified Gaussian Renormalized Fluctuation theory to investigate the counter-intuitive phenomenon of repulsion between two oppositely charged surfaces and discuss its relationship with overcharging. By accurately accounting for the effect of spatially varying ion-ion correlations, we capture these repulsive forces for divalent, trivalent as well as tetravalent ions, in quantitative agreement with reported simulation results. We show that the opposite-charge repulsion is long-ranged with an effective length scale of a few nanometers. The strength of opposite-charge repulsion increases monotonically with the multivalent salt concentration, in stark contrast with the non-monotonic salt concentration dependence of other ion correlation-driven phenomena, such as overcharging and like-charge attraction. We also elucidate that the origin of the opposite-charge repulsion is the large amount of ions attracted to the double layer as a result of ion-ion correlations, leading to higher osmotic pressure and stronger screening of the electrostatic attraction, which results in an overall repulsive force between two
oppositely charged surfaces. Furthermore, we demonstrate that there is no causal relationship between opposite-charge repulsion and the overcharging of the surface. Opposite-charge repulsion is accompanied by overcharging at large separation distances, but can also occur in normal double layers without overcharging at intermediate separation distances. 
\end{abstract}

\maketitle

\section{\label{sec:level1}Introduction\protect\\}

The electrostatic force between two charged surfaces in the presence of a salt solution is the key to determining equilibrium and non-equilibrium properties in a wide variety of systems related to electrochemistry\cite{Bukosky2015SimultaneousFields,Woehl2015BifurcationMinimum,Ajdari2000PumpingArrays,Itskovich1977ElectricCharacteristic}, materials science\cite{israelachvili2011intermolecular,Larsen1997Like-chargeCrystallites,Dinsmore1998Self-assemblyCrystals}, and biology\cite{Caccamo_2000,Butler2003IonAttraction, Zhang2008ReentrantCounterions,Nassar2021TheTheory,Cevc1990MembraneElectrostatics,Kozlov2014MechanismsMembranes}. In the classical Derjaguin-Landau-Verwey-Overbeek (DLVO) framework\cite{israelachvili2011intermolecular}, this electrostatic force is described by the mean-field Poisson-Boltzmann (PB) theory, which is long-ranged compared to the short-range Van der Waals attraction. This mean-field PB theory does not account for ion-ion correlations, variations in dielectric permittivity as well as excluded volume effect of ions and solvent molecules. Consistent with physical intuition, mean-field PB predicts a repulsive force between two like-charged surfaces and an attractive force between two oppositely charged surfaces at all separation distances. However, numerous experimental and simulation studies have reported attraction between like-charged surfaces\cite{Kekicheff1993ChargeElectrolyte,Zohar2006ShortSurfaces,Kumar2017InteractionsIons,linselca1999,Linse2000,Wu1999MonteSalts,Angelescu2003MonteAdded,Zhang2016PotentialEffect} and repulsion between oppositely charged surfaces\cite{Besteman2004DirectPhenomenon,Besteman2005ChargeDensity,Trulsson2006RepulsionElectrolytes,Trulsson2007RepulsionParticles,Popa2010ImportanceSurfaces,MontesRuiz-Cabello2014AccurateTheory,AntilaInteractionRods,Antila2017RepulsionConfinement,Moazzami-Gudarzi2018CiteThis,Lin2019ApparentSolutions} in the presence of multivalent salts. The origin of these counter-intuitive phenomena is the electrostatic correlation between ions, the strength of which dominates thermal forces in the case of multivalent ions\cite{Trulsson2006RepulsionElectrolytes,Linse2000}. Since the mean-field PB does not account for ion-ion correlations, it fails to even qualitatively capture the aforementioned phenomena. While like-charge attraction has been extensively studied by theories beyond mean-field PB\cite{Patey1980TheApproximation,Kjellander1988Double-LayerSwelling,Rouzina1996MacroionCloud, Chu1996AttractiveParticles,Ha1997Counterion-MediatedRods,Arenzon1999SimplePolyions,Diehl1999Density-functionalPlates, Grosberg2002Colloquium:Systems,Kudlay2003PrecipitationSolutions,Netz2000BeyondFunctions,Buyukdagli2017Like-chargeMolecules,Suematsu2018CiteAs, Misra2019TheoryElectrolytes, Chen2020MultivalentPolyelectrolytes,gupta_prl_overcharging,gupta_force}, opposite-charge repulsion has received considerably less attention\cite{Hatlo2009ACouplings,ShiqiZhou2019EffectiveLike}. Understanding opposite-charge repulsion is of great importance, as the interaction between oppositely-charged surfaces is at the core of numerous practical applications like cement formation\cite{Ferrari2010InteractionMeasurements}, papermaking\cite{Varennes1988EffectsFlow}, food processing\cite{Guzey2006FormationIndustry}, and surface patterning\cite{Agheli2006LargeApplications}, as well as biological processes such as protein binding\cite{Aranda-Espinoza1999ElectrostaticObjects,chromatinphysics,Pieters2015NaturalAssemblies} and targeted drug delivery\cite{Gelbart2007DNAInspiredElectrostatics,Solis2007FlexibleCounterattract}.\par 

The force measurements by Besteman et al.\cite{Besteman2004DirectPhenomenon,Besteman2005ChargeDensity} provided one of the very first experimental evidence that opposite-charge repulsion is an outcome of ion-ion correlations. They employed an atomic force microscope (AFM) setup to measure the force between a positively-charged amine-terminated surface and  a silica bead which originally carries negative charges. It was found that the force between these two surfaces changes from attractive to repulsive at a critical multivalent salt concentration. This critical concentration strongly depends on the ion valency and weakly depends on the chemical identity of the ions. A similar result has also been observed from the AFM measurements conducted by Borkovec and coworkers\cite{Popa2010ImportanceSurfaces,MontesRuiz-Cabello2014AccurateTheory,Moazzami-Gudarzi2018CiteThis}. Using Monte-Carlo simulations for the primitive model of electrolytes, Trulsson et al.\cite{Trulsson2006RepulsionElectrolytes,Trulsson2007RepulsionParticles} were able to reproduce these repulsive force profiles, thus confirming the purely electrostatic origin of opposite-charge repulsion. It was found that opposite-charge repulsion is long-ranged, acting up to a length scale of a few nanometers. This is quite different from ion correlation-driven like-charge attraction, which occurs at the length scale of a few angstroms\cite{Linse2000,agrawal2023lca}.\par

The existing studies have indicated that opposite-charge repulsion is related to another ion-correlation induced phenomenon known as overcharging\cite{Grosberg2002Colloquium:Systems,Agrawal2022OnLayers}, which is defined as the excess accumulation of counterions near a charged surface. However, the relationship between these two phenomena remains unclear. Trulsson et al.\cite{Trulsson2006RepulsionElectrolytes} reported that repulsion is accompanied by overcharging at large separations between the two surfaces, but for the same salt concentration at intermediate separations, repulsion can occur without overcharging. These observations suggest that overcharging is not the cause of opposite-charge repulsion. While the simulations of Trulsson et al. were limited to low salt concentrations, it is natural to ask whether this conclusion holds for high salt concentrations as well. Particularly, does the strength of repulsion increase monotonically with salt concentration? This is an important question to ask as the strength of other correlation-induced phenomena like overcharging and like-charge attraction have been shown to exhibit non-monotonic salt concentration dependence\cite{Wu1999MonteSalts,agrawal2023lca,agrawal_aiche,VanDerHeyden2006ChargeCurrents,Hsiao2006Salt-inducedPolyelectrolytes}. Furthermore, it is also desirable to understand how the repulsive force changes with the addition of a a secondary monovalent salt. This is crucial as multivalent and monovalent salt mixtures widely exist in many biological and geological systems\cite{Levin2002ElectrostaticBiology}.\par 
On the theoretical side, Hatlo and Lue\cite{Hatlo2009ACouplings} developed a variational approach in the field theoretical framework to model opposite-charge repulsion and overcharging in multivalent salt solutions. Their theory used a point-charge model for the ions, not only overestimating the ion correlations but also ignoring the excluded volume effect of both ions and solvent molecules. They also noted that an arbitrary mathematical operator needs to be introduced to decouple of the short and long-range components of the ion correlations. Although they were able to show good agreement with the simulation results of Trulsson et al.\cite{Trulsson2006RepulsionElectrolytes} for divalent and trivalent salts, their method failed to predict any repulsion for tetravalent salts, which is inconsistent with the simulation results. Recently, Zhou\cite{ShiqiZhou2019EffectiveLike} employed a classical density functional theory (DFT)-based approach and included the contribution of ion correlations to the free energy using second-order perturbation around the bulk density of ions. This perturbative approach inherently limits the applicability of the method to systems where ion concentrations at the surface are close to the bulk salt concentration. This assumption is not valid if the surface charge density or ion valency is high. Furthermore, DFT theories are computationally challenging and often use density weighting functions that are specific to a particular geometry or system, again limiting its generalization.\par

An accurate treatment of the spatially varying ion correlations on the structure of overlapping double layers and their free energy is necessary to understand the mechanism behind long-range opposite-charge repulsion. In our previous work, we have developed a modified Gaussian Renormalized Fluctuation theory and demonstrated the effectiveness of the theory in capturing such inhomogeneous ion correlations by successfully modeling the vapor-liquid interface in ionic fluids\cite{Agrawal2022Self-ConsistentFluids}, overcharging and charge inversion in nanochannels\cite{Agrawal2022OnLayers,agrawal_aiche}, and like-charge attraction between two surfaces\cite{agrawal2023lca}.
In this work, we use this theory to study the force between two approaching oppositely charged surfaces immersed in a multivalent salt solution. The force profiles calculated by our theory are in quantitative agreement with the simulation results by Trulsson et al.\cite{Trulsson2006RepulsionElectrolytes} for divalent, trivalent, as well as tetravalent ions. We examine the nature of the repulsive force in the entire salt concentration regime and elucidate the relationship between overcharging and opposite-charge repulsion. Furthermore, we investigate how the force change with the addition of monovalent salt to a multivalent salt solution.

\section{Theory}
\label{sec:theory}

We consider a system of two oppositely charged plates, one located at $z=0$ with a uniform negative surface charge density $\sigma_1$ and the other located at $z=D$ with a uniform positive surface charge density $\sigma_2$. The two plates are immersed in a solution with $n$ type of salts. Each salt $i$ ($i$ ranges from 1 to $n$) has cations of valency $q_{i,+}$ and anions of valency $q_{i,-}$. The electrolyte solution between the two plates is connected to a bulk reservoir of salt concentrations $c_{i,\mathrm{b}}$. The dielectric permittivity at a position $\mathbf r$ is given by the function $\varepsilon ({\mathbf r})$. Instead of the point-charge model for mobile ions as used in the original variational formulation of the Gaussian Fluctuation theory by Netz and Orland\cite{Netz2003VariationalSystems}, we consider ions to be finite-sized and the ionic charge to have a finite spread given by a distribution function $h_{i,\pm}(\mathbf{r'}-\mathbf{r})$, where $\mathbf{r}$ is the position of the center of the ion. This charge spread model is introduced to avoid the overestimation of ion-ion correlation aroused by the point-charge model and remove any divergence issues from the fluctuation contribution thus acting as a renormalization factor\cite{Wang2010FluctuationEnergy}. We also include the excluded volume effect of ions and solvent molecules by taking their volume as $v_{i,\mathrm{\pm/s}}$. For this system, the modified Gaussian Renormalized Fluctuation theory derived in our previous work leads to the following set of self-consistent equations for the electrostatic potential $\psi(z)$, ion concentrations $c_{i,\pm}(z)$, self-energy of ions $u_{i,\pm}(z)$, and electrostatic correlation function $G(\mathbf{r'},\mathbf{r''})$: 

\begin{equation}
\begin{split}
{-\nabla.[\epsilon(z)\nabla\psi(z)]} &= \sigma_1\delta(z) + \sigma_2\delta(z-h)  + \\ & \quad \sum_{i=1}^{n} \left[q_{i,+}{c_{i,+}(z)} - q_{i,-}{c_{i,-}(z)}\right]
\label{eq:psi}
\end{split}
\end{equation}
\begin{eqnarray}
{c_{i,\pm}(z)}= \frac{ \mathrm{e}^{\mu_{i,\pm}}}{v_{i,\pm}}\exp[\mp q_{i,\pm}\psi(z) - u_{i,\pm}(z) -v_{i,\pm}\eta(z)]
\label{eq:conc}
\end{eqnarray}
\begin{eqnarray}
{\textit{u}_{i,\pm}(\mathbf{r})}=\frac{q_{i,\pm}^2}{2}\int d\mathbf{r}'d\mathbf{r}'^{\prime}h_{i,\pm}(\mathbf{r'},\mathbf{r})G(\mathbf{r'},\mathbf{r'^{\prime}})h_{i,\pm}(\mathbf{r'^{\prime}},\mathbf{r})
\label{eq:selfe}
\end{eqnarray} 
\begin{eqnarray}
{-\nabla_{\mathbf{r'}}.[\epsilon(\mathbf{r'})\nabla_{\mathbf{r'}}G(\mathbf{r'},\mathbf{r''})]} + 2I(\mathbf{r'})G(\mathbf{r'},\mathbf{r''}) = \delta(\mathbf{r'}-\mathbf{r''})
\label{eq:greens}
\end{eqnarray}
where $2I(\mathbf{r'})= \epsilon(\mathbf{r'})\kappa^2(\mathbf{r'}) = \sum_{i=1}^{n}\left[c_{i,+}(\mathbf{r'})q_{i,+}^2 + q_{i,-}^2c_{i,-}(\mathbf{r'})\right]$, $\epsilon({\mathbf{r'}})=kT\varepsilon_{0}\varepsilon ({\mathbf r'})/e^2$ is the scaled permittivity with $\varepsilon_{0}$ as the vacuum permittivity and $e$ as the elementary charge. $\eta(z)$ is the field accounting for the excluded volume effect via enforcing the incompressibility constraint as: 
\begin{equation}
\eta(z) =  -\frac{1}{v_s}{\ln(1 - \sum_{i=1}^{n} \left[v_{i,+}c_{i,+}(z) +v_{i,-} c_{i,-}(z)\right])}
\label{eq:eta}
\end{equation}  
$\mu_{i,\pm}$ are chemical potentials of ions which are determined from the bulk salt concentration $c_{i,\mathrm{b}}$ through
\begin{eqnarray}
\mu_{i,\mathrm{\pm}}= u_{i,\mathrm{\pm,b}} + \ln{c_{i,\mathrm{\pm,b}}v_{i,\mathrm{\pm}}}  +v_{i,\mathrm{\pm}}\eta_\mathrm{b}
\label{eq:mu_ions}
\end{eqnarray}
$u_{i,\mathrm{\pm,b}}$ being the self-energy of ions in the bulk, where the electrostatic potential is zero and the ion distribution is uniform. 

The grand free energy ${W}$ of this system is given by
\begin{equation}
\begin{split}
W & = \int d{z}\frac{\psi}{2}\left(\sigma_1\delta(z) + \sigma_2\delta(z-h) -\sum_{i=1}^{n} \left[q_{i,+}{c_{i,+}} - q_{i,-}{c_{i,-}}\right]\right)
\\ & \quad - \int d{z}\left[\sum_{i=1}^{n} {c_{i,+}} + {c_{i,-}}\right]
\\ & \quad - \int d{z}\frac{(1 - \sum_{i=1}^{n} \left[v_{i,+}c_{i,+}(z) +v_{i,-} c_{i,-}(z)\right])}{v_s} 
\\ & \quad + \int d{z}\left[\frac{\ln(1 - \sum_{i=1}^{n} \left[v_{i,+}c_{i,+}(z) +v_{i,-} c_{i,-}(z)\right])}{v_s}\right]
\\  &\quad + \frac{1}{2}\int d{z}\int d{\bf{r'}}\int d{\bf{r'^{\prime}}}\int_{0}^{1}d\tau[G({\bf{r'}},{\bf{r'^{\prime}}},\tau)  \\ & \quad - G({\bf{r'}},{\bf{r'^{\prime}}})]\left[\sum_{i=1}^{n}\sum_{j=+,-}q_{i,j}^2c_{i,j}(z) h_{i,j}({\bf{r'}}-{\bf{r}})h_{i,j}({{\bf{r'^{\prime}}}-\bf{r}})\right]
\end{split}
\label{eq:minfe}
\end{equation}
The first and second lines in Eq. \ref{eq:minfe} retain the free energy form as obtained from the mean-field PB. It should be noted that the fluctuation effect is implicitly included in $\psi$ and $c_{i,\pm}$ through the feedback from the self-energy of ions. The third and fourth lines are the translational entropy of the solvent molecules coming from the inclusion of the excluded volume effect. The fifth and sixth lines are the charging term explicitly accounting for the contribution of the electrostatic fluctuation to the free energy, with $\tau$ the charging variable\cite{Wang2015OnSurfaces}. The pressure between the two plates can be calculated from the above free energy expression as: 
\begin{eqnarray}
P = -\left(\frac{\partial W}{\partial h}\right)_{\mu_\mathrm{\pm}} - P_\mathrm{b}
\label{eq:force}
\end{eqnarray}
where $P_\mathrm{b}$ is the osmotic pressure of the bulk reservoir containing salts at a uniform concentration of $c_{i,\mathrm{b}}$. \par 

The inclusion of the finite-charge spread model introduces an additional complexity of dealing with the dual-length scale problem associated with the numerical calculation of the spatially varying self-energy of ions. To accurately calculate $u$ we need to evaluate $G$ at both the length scale of the ion and that of the separation between the two surfaces. Opposite-charge repulsion is known to be long-ranged acting at the length scale of a few nanometers, while the length scale of the ion is around 1-4 \AA. The separation distances for which we want to calculate the double-layer structure and properties are hence one order of magnitude larger than the ion size. An impractically large number of grid points in the numerical calculation is needed if $G$ is resolved at these two very different length scales simultaneously, thus making the problem intractable. To bypass this dual-length scale issue, we apply the decomposition scheme for $G$ developed in our previous work\cite{Agrawal2022Self-ConsistentFluids}, where $G$ is decomposed into a short-range contribution $G_\mathrm{S}$ and long-range contribution $G_\mathrm{L}$ as: 
\begin{equation}
G(\mathbf{r'},\mathbf{r''}) = G_\mathrm{S}(\mathbf{r'},\mathbf{r''},\mathbf{r}) + G_\mathrm{L}(\mathbf{r'},\mathbf{r''},\mathbf{r})
\label{eq:greens_dec}
\end{equation}
The motivation behind the above decomposition is to decouple the short-range contribution associated with the local electrostatic environment and the long-range contribution associated with spatially varying ionic strength and dielectric permittivity through $G_\mathrm{S}$ and $G_\mathrm{L}$, respectively. Since $G_\mathrm{S}$ accounts for local correlations, its equation is constructed from Eq. \ref{eq:greens} using the local ionic strength $I(\mathbf{r})$ and dielectric permittivity $\epsilon(\mathbf{r})$ instead of spatially varying $I(\mathbf{r'})$ and $\epsilon(\mathbf{r'})$ as:
\begin{equation}
-\epsilon(\mathbf{r}){\nabla_{\mathbf{r'}}^2G_\mathrm{S}(\mathbf{r'},\mathbf{r'^{\prime}},\mathbf{r})} + 2I(\mathbf{r})G_\mathrm{S}(\mathbf{r'},\mathbf{r'^{\prime}},\mathbf{r}) = \delta(\mathbf{r'}-\mathbf{r'^{\prime}})
\label{eq:greens_short}
\end{equation} 
The above equation has a Debye-H\"{u}ckel style analytical solution
\begin{equation}
G_\mathrm{S}(\mathbf{r'},\mathbf{r'^{\prime}},\mathbf{r})  =  \frac{\mathrm{e}^{-\kappa(\mathbf{r})|\mathbf{r'}-\mathbf{r'^{\prime}}|}}{4\pi\epsilon(\mathbf{r})|\mathbf{r'}-\mathbf{r'^{\prime}}|}
\label{eq:gsol_short}
\end{equation} 
Following Eq. \ref{eq:greens_dec}, $G_\mathrm{L}$ can be obtained by subtracting Eq. \ref{eq:greens_short} from Eq. \ref{eq:greens} as: 
\begin{equation}
{-\nabla_{\mathbf{r'}}.[\epsilon(\mathbf{r'})\nabla_{\mathbf{r'}}G_\mathrm{L}(\mathbf{r'},\mathbf{r'^{\prime}},\mathbf{r})]} + 2I(\mathbf{r'})G_\mathrm{L}(\mathbf{r'},\mathbf{r'^{\prime}},\mathbf{r})= S(\mathbf{r'},\mathbf{r'^{\prime}},\mathbf{r})
\label{eq:greens_long}
\end{equation}
where the non-local source term $S$ is
\begin{equation}
\begin{split}
S(\mathbf{r'},\mathbf{r'^{\prime}},\mathbf{r}) & = \nabla_{\mathbf{r'}}.((\epsilon(\mathbf{r'})-\epsilon(\mathbf{r}))\nabla_{\mathbf{r'}}G_\mathrm{S}(\mathbf{r'},\mathbf{r'^{\prime}},\mathbf{r}))\\ & \quad - 2(I(\mathbf{r'}) - I(\mathbf{r}))G_\mathrm{S}(\mathbf{r'},\mathbf{r'^{\prime}},\mathbf{r}) 
\end{split}
\end{equation}
Next, although the charge distribution function $h$ is a crucial factor in calculating the contribution of the local electrostatic environment to self-energy, the evaluation of the long-range component of self-energy associated with $G_\mathrm{L}$ is not sensitive to the exact form of $h$ since the width of the double layer is much larger than the ion size. As a result, we can take the point charge limit when evaluating the long-range component of self-energy\cite{Agrawal2022Self-ConsistentFluids}, reducing Eq. \ref{eq:selfe} to    
\begin{eqnarray}
\textit{u}_{i,\pm}(\mathbf{r})=  \frac{q_{i,\pm}^2}{2}\int _{\mathbf{r}',\mathbf{r}'^{\prime}}h_{i,\pm}G_\mathrm{S}h_{i,\pm} + \frac{q_{i,\pm}^2}{2}G_\mathrm{L}(\mathbf{r},\mathbf{r},\mathbf{r})
\label{eq:selfe_hybrid}
\end{eqnarray} \par

$h_{i,\pm}(\mathbf{r'}-\mathbf{r})$ in general can be any arbitrary function as long as the Born solvation energy is retained. Here, we adopt the following Gaussian form for mathematical convenience 
\begin{equation}
{h_{i,\pm}(\mathbf{r'} -\mathbf{r})}  =  \left({\frac{1}{2a_{i,\pm}}}\right)^{3/2}\exp\left[\frac{-\pi{}(\mathbf{r'}-\mathbf{r})^{2}}{2a^2_{i,\pm}}\right]
\label{eq:cspread}
\end{equation}
where $a_{i,\pm}$ is the Born radius of the ion. For the above choice of $h$, the short-range component of the self energy $u_{\pm,S}$ (the first term on the right hand side of Eq. \ref{eq:selfe_hybrid}),  becomes\cite{Agrawal2022Self-ConsistentFluids} 
\begin{equation}
\begin{split}
u_{i,\mathrm{\pm,S}}(\mathbf{r}) & = \frac{q^2_{i,\pm}}{8\pi\epsilon(\mathbf{r})a_{i,\pm}} -\frac{q^2_{i,\pm} \kappa(\mathbf{r})}{8\pi\epsilon(\mathbf{r})} \\ & \quad \times\exp\left(\frac{{a_{i,\pm}^2\kappa(\mathbf{r})}^2}{\pi}\right)\mathrm{erfc}\left(\frac{{a_{i,\pm}\kappa(\mathbf{r})}}{\sqrt{\pi}}\right)
\label{eq:selfe_short}
\end{split}
\end{equation}
\begin{figure*}
\captionsetup[subfigure]{labelformat=empty}
    \begin{subfigure}{0.68\columnwidth}
    \includegraphics[width=\columnwidth]{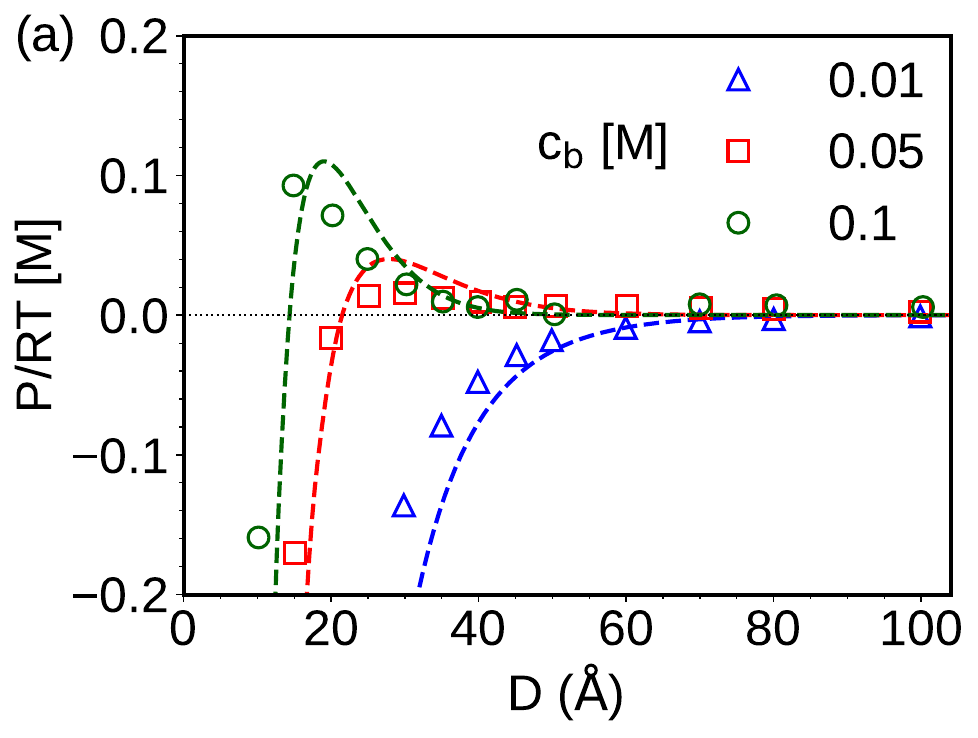}
    \caption{}
    \label{fig:div_p}
    \end{subfigure}  
    \begin{subfigure}{0.68\columnwidth}
        \includegraphics[width=\columnwidth]{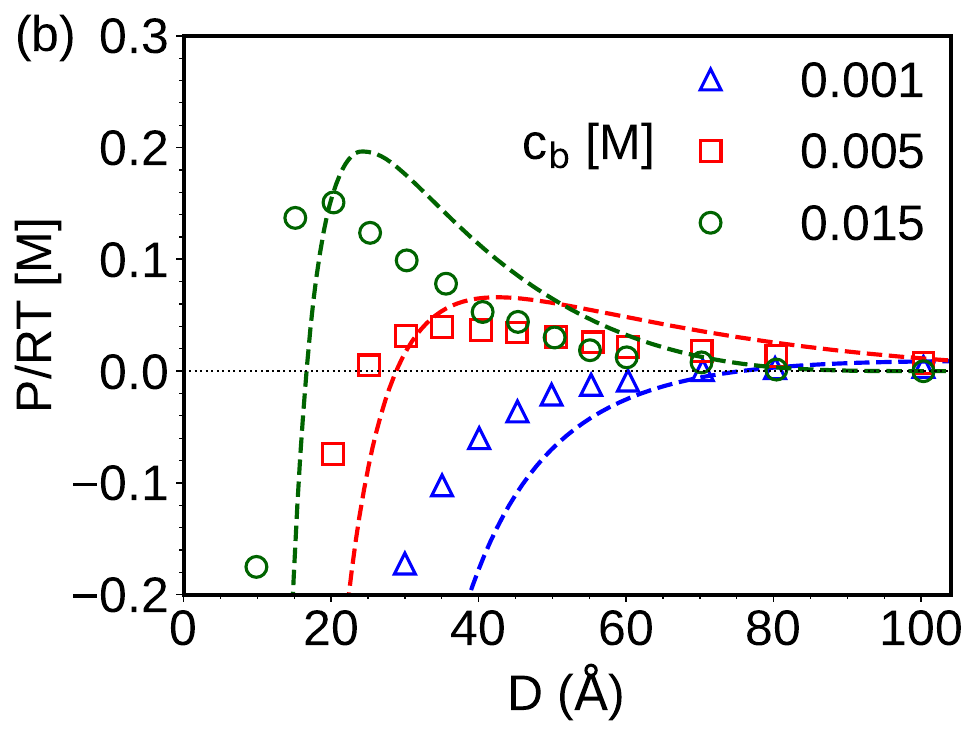}
        \caption{}
        \label{fig:tri_p}
    \end{subfigure} 
    \begin{subfigure}{0.68\columnwidth}
        \includegraphics[width=\columnwidth]{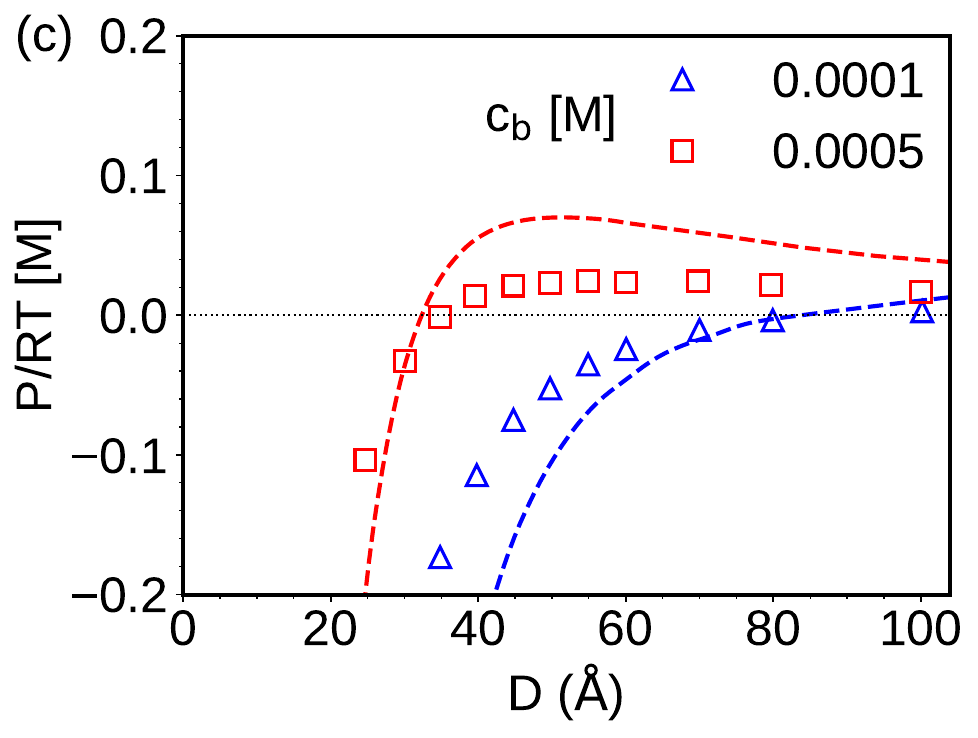}
        \caption{}
        \label{fig:tetra_p}
    \end{subfigure}
\caption{Transition from pure attraction to long-range repulsion in overlapping double layers as the concentration of multivalent salt c$_\mathrm{b}$ increases. The net pressure $P$ between two surfaces is plot as a function of separation distance $D$ for a) divalent $q_+$ = 2, b) trivalent $q_+$ = 3 and  c) tetravalent $q_+$ = 4 cations. Dashed lines represent results calculated by our theory and symbols represent simulation data from Trulsson et al.\cite{Trulsson2006RepulsionElectrolytes} for the comparison. $\sigma_1$ = -0.3204 C/m$^2$, $\sigma_2$ = 0.1602 C/m$^2$, $q_-$ = 1, and $a_{-} = 1.5$ \AA. The Born radius $a_\mathrm{+}$ are chosen to be 1.6 \AA, 2.5 \AA, and 2.5 \AA\ for divalent, trivalent, and tetravalent cations, respectively, for the best fit. }
\label{fig:val_panel}
\end{figure*}

The long range component of the self-energy is the same-point $G_\mathrm{L}$. Interested readers are referred to the our previous work\cite{Agrawal2022Self-ConsistentFluids,Agrawal2022OnLayers,agrawal_aiche} for the numerical scheme to solve $G_\mathrm{L}$. The iteration scheme for solving Eq. \ref{eq:psi} and \ref{eq:conc} was partially adopted from Xu and Maggs\cite{Xu2014SolvingEquations}. The Python code for solving the equations in this work was written on top of open-source spectral methods based differential equation solver \textit{Dedalus}, developed by Burns et al.\cite{burns_dedalus}

\section{Results and Discussion}
\label{sec:Results}

In this work, we study the force between two plates mediated by pure multivalent salt solutions as well as their mixtures with monovalent salt. The surface charge density of the two plates are $\sigma_1 = -0.3204$  C/m$^2$ and $\sigma_2 = 0.1602$ C/m$^2$, respectively. We use the primitive model for electrolytes where the dielectric constant is assumed to be uniform throughout the system and is taken to be the value of water at 298 K, i.e., $\varepsilon(\bm{r}) = 78.7$. The values of surface charge density and dielectric permittivity model adopted here is the same as those used in the work of Trulsson et al.\cite{Trulsson2006RepulsionElectrolytes}, which facilitates a direct comparison with their simulation results. For simplicity, the excluded volume of ions and solvent molecules are set to be same, $v_{i, \pm}=v_{s}$. \par

Figure \ref{fig:val_panel} plots the pressure $P$ as a function of the separation distance $D$ between the two surfaces immersed in divalent, trivalent and tetravalent salt solutions. Our theoretical predictions are in very good quantitative agreement with the simulation results of Trulsson et al.\cite{Trulsson2006RepulsionElectrolytes} for all the different cation valencies. At low salt concentrations, the ion correlation effect is not significant, the force is universally attractive at all separation distances. This is consistent with the well-known mean-field PB results of the interactions between two oppositely charged objectives. However, the effect of ion correlations becomes more pronounced as salt concentration increases. The long-range force turns gradually from attraction to repulsion, leading to the emergence of the long-range opposite-charge repulsion. It is worth noting that the force remains attractive at short ranges even in the presence of long-range opposite-charge repulsion. This duality of long-range repulsion and short-range attraction gives rise to a maxima in the pressure curve, which is also in agreement with independent force measurements by Besteman et al.\cite{Besteman2004DirectPhenomenon, Besteman2005ChargeDensity} and Borkovec et al.\cite{Popa2010ImportanceSurfaces,MontesRuiz-Cabello2014AccurateTheory,Moazzami-Gudarzi2018CiteThis}. With only multivalent ion size as an adjustable parameter, we are able to quantitatively capture the pressure curves at both short and long ranges. To our knowledge, this is the first theoretical work in the literature to quantitatively capture these pressure curves for all cation valencies for the entire salt concentration range. \par     
\begin{figure}
\includegraphics[width=\columnwidth]{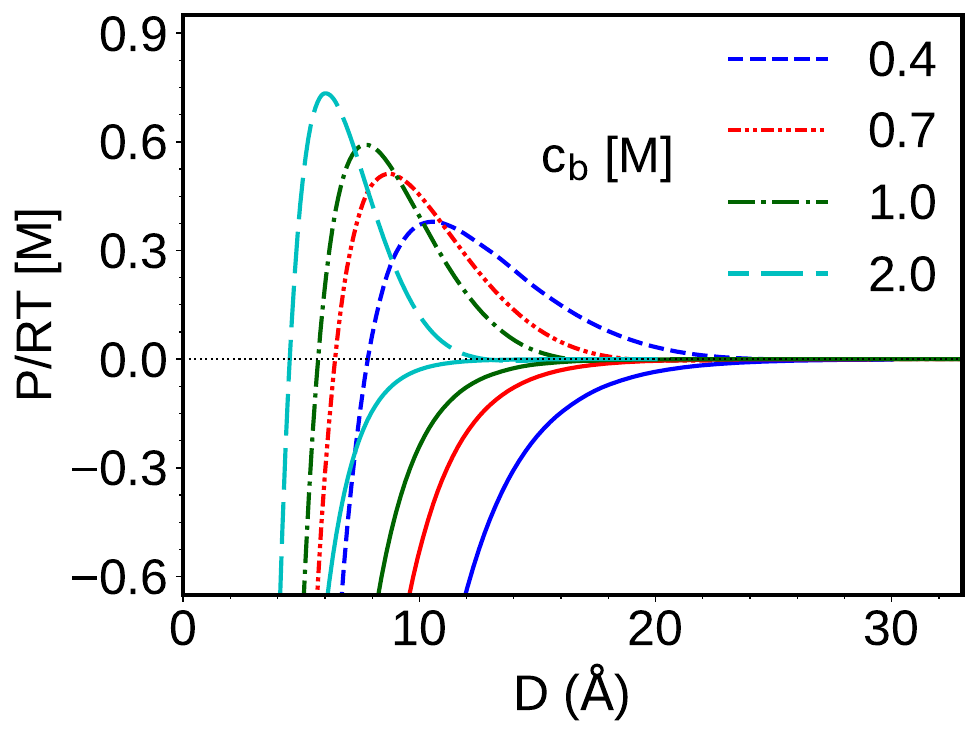}
  \caption{Monotonic increase of the net repulsive pressure between two surfaces as a function of separation distance $D$ with different divalent salt concentrations c$_\mathrm{b}$. The results predicted by our theory are presented by dashed lines in comparison with solid lines representing the mean-field PB results. $\sigma_1$ = -0.3204 C/m$^2$, $\sigma_2$ = 0.1602 C/m$^2$, $q_+$ = 2, $q_-$ = 1, and $a_\mathrm{+,s} = 1.6$ \AA.}
  \label{fig:force_conc}
\end{figure}
\begin{figure*}
\captionsetup[subfigure]{labelformat=empty}
    \begin{subfigure}{0.68\columnwidth}
    \includegraphics[width=\columnwidth]{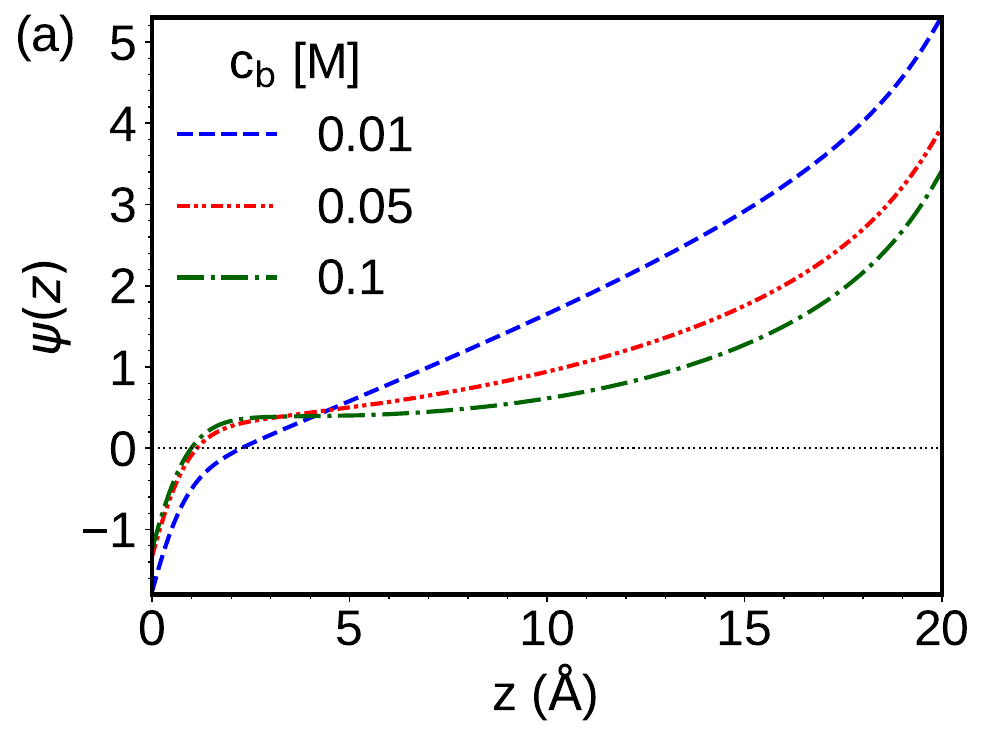}
    \caption{}
    \label{fig:psi}
    \end{subfigure}  
    \begin{subfigure}{0.68\columnwidth}
        \includegraphics[width=\columnwidth]{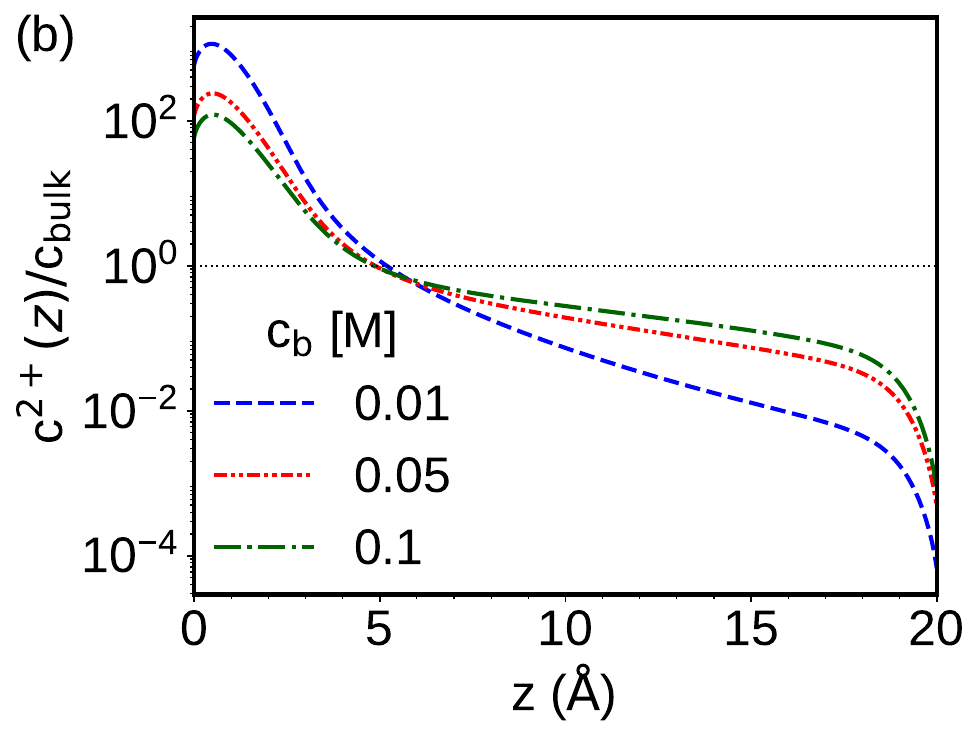}
        \caption{}
        \label{fig:counterion}
    \end{subfigure} 
    \begin{subfigure}{0.68\columnwidth}
        \includegraphics[width=\columnwidth]{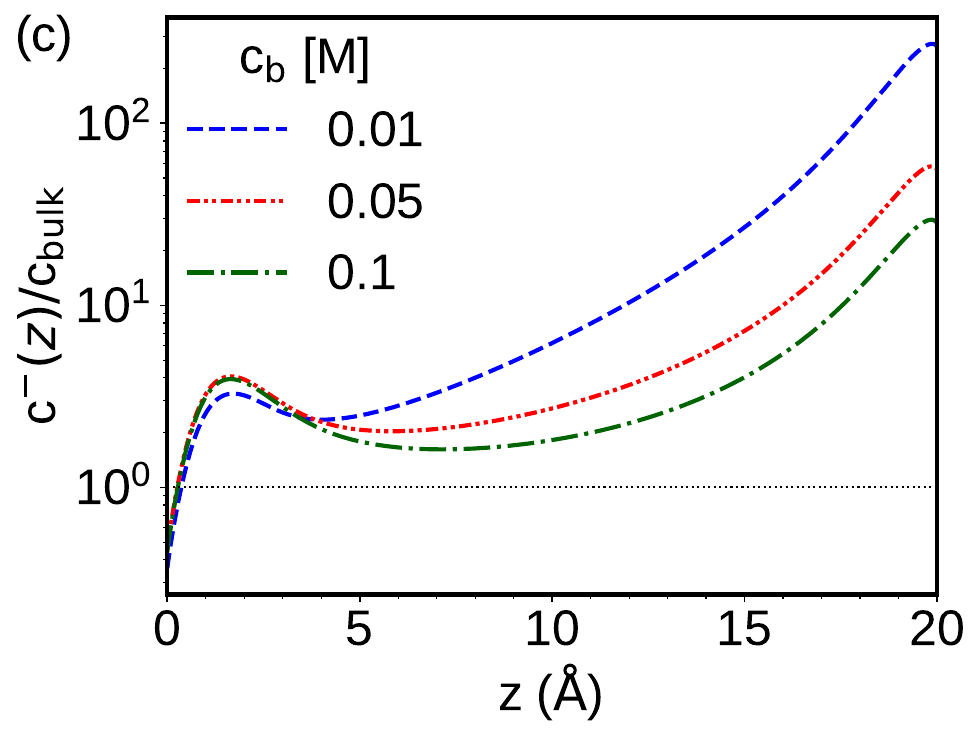}
        \caption{}
        \label{fig:coion}
    \end{subfigure}
\caption{Structure of the electrical double layer as divalent salt concentration increases. a) Electrostatic potential profile $\psi(z)$, b) Counterion concentration $c^{2+}$(z) profile with respect to bulk counterion concentration $c^{2+}_\mathrm{bulk}$, and c) b) Coion concentration $c^{-}$(z) profile with respect to bulk counterion concentration $c^{-}_\mathrm{bulk}$. $\sigma_1$ = -0.3204 C/m$^2$, $\sigma_2$ = 0.1602 C/m$^2$, $D = 20$ \AA,  $q_+$ = 2, $q_-$ = 1, $a_{+} = 1.6$ \AA\ and $a_{-} = 1.5$ \AA.}
\label{fig:edl_str}
\end{figure*}
It has been reported that other ion correlation-induced phenomena such as overcharging and like-charge attraction show non-monotonic dependence on salt concentration, whose strengths are enhanced at medium salt concentrations but suppressed at high salt concentrations. To understand the nature of long-range opposite-charge repulsion, it is necessary to investigate its trend at high salt concentrations. Figure \ref{fig:force_conc} plots pressure profiles predicted by our theory for pure divalent salt up to $c_\mathrm{b} = 2.0$ M, in comparison with the mean-field PB results. The force predicted by the mean-field PB theory is always attractive at all salt concentrations, and gets suppressed as $c_\mathrm{b}$ increases. In stark contrast, our theory with a systematic inclusion of the ion correlations predicts long-range repulsion in the medium to high salt concentration regime. Most importantly, the strength of repulsion is found to keep increasing as salt concentration increases, with a shift of the maxima to a smaller $D$. Although opposite-charge repulsion and like-charge attraction are both caused by electrostatic correlation, our results clearly demonstrate that these two anomalous phenomena are significantly different: they are not only of different effective range, but also have different salt concentration dependence. To be specific, like-charge attraction is short-ranged (acting at a distance of few angstroms) and depends non-monotonically on multivalent salt concentration. In contrast, opposite charge repulsion is long-ranged (acting at a distance of few nanometers) and increases monotonically with multivalent salt concentration.\par

\begin{figure}
\includegraphics[width=\columnwidth]{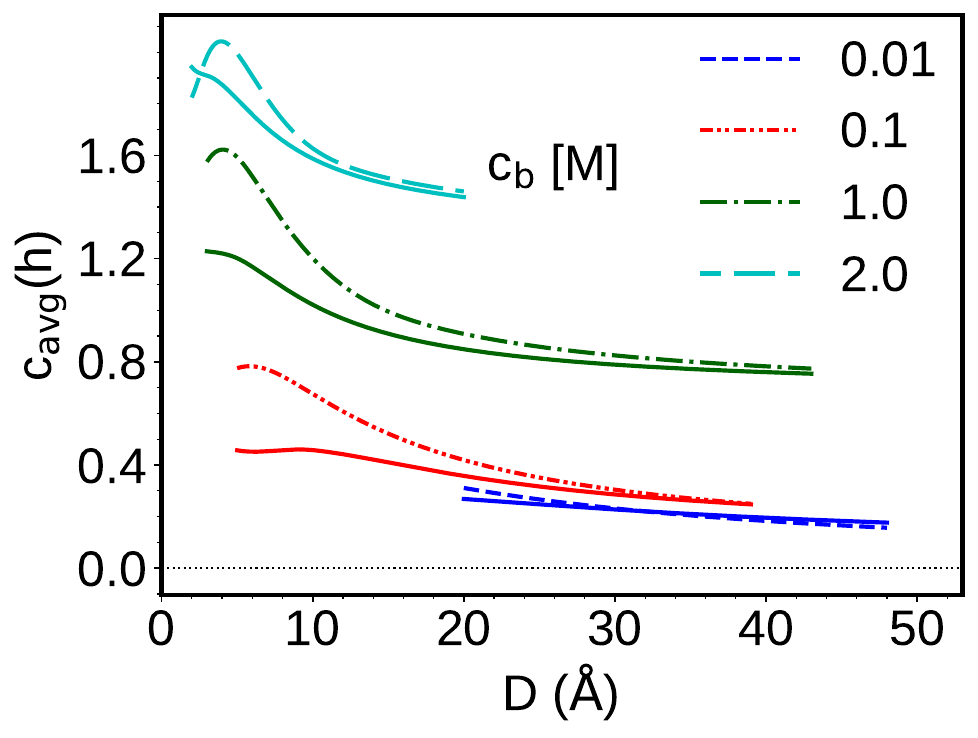}
  \caption{Average ion concentrations in the slit between the two plates $c_{avg}$. The results predicted by our theory are presented by dashed lines in comparison with solid lines representing the mean-field PB results. $\sigma_1$ = -0.3204 C/m$^2$, $\sigma_2$ = 0.1602 C/m$^2$, $q_+$ = 2, $q_-$ = 1, and $a_\mathrm{+,s} = 1.6$ \AA.}
  \label{fig:conc_conc_ocr}
\end{figure}

In Figure \ref{fig:edl_str} we show the molecular picture of electrical double layers corresponding to the three force curves for the case of divalent salt solution presented in Figure \ref{fig:div_p}. Although the force between the two plates transitions from attractive ($c_b = 0.01$ M) to repulsive ($c_b = 0.05$ M and 0.1 M), there is no significant change in the nature of the $\psi(z)$ and $c(z)$ profiles, suggesting a nontrivial origin for the repulsive force. Therefore, to uncover the origin of opposite-charge attraction, we analyze different components in the free energy. The free energy of the system can be divided into two major components: a repulsive entropic component due to the ionic osmotic pressure and an attractive electrostatic component due to the interactions of the oppositely charged surfaces. The more ions coming to the slit between the two plates, the stronger the osmotic repulsion and the weaker the electrostatic attraction due to the enhanced screening. Figure \ref{fig:conc_conc_ocr} quantifies the average ion concentration $c_{avg}$ in the slit as a function of $D$. Compared to the mean-field PB results, significant additional amount of ions are attracted to the slit as a result of ion correlations. This paves the way for the dominance of the entropic component in the free energy in a certain regime of $D$, turning the force from attraction to repulsion. The locations of the maximum of $c_{avg}$ are close to the peak positions of $P$ in Figure \ref{fig:force_conc}, which confirms that the entropic component is responsible for the opposite-charge repulsion.\par

\begin{figure*}
\captionsetup[subfigure]{labelformat=empty}
    \begin{subfigure}{\columnwidth}
    \includegraphics[width=\columnwidth]{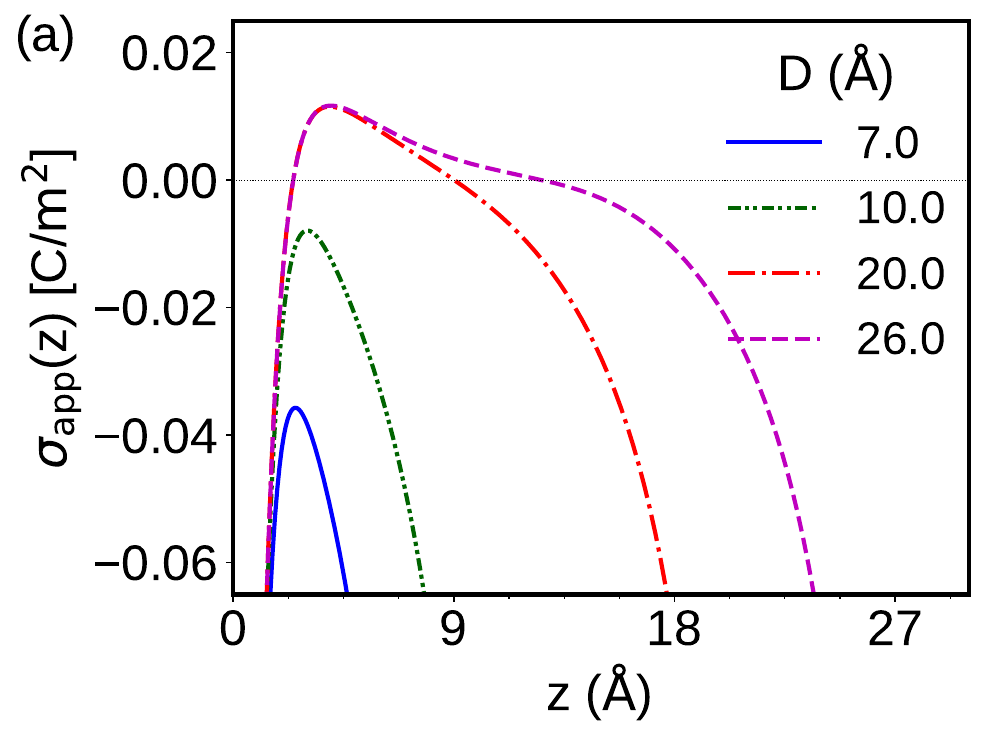}
    \caption{}
    \label{fig:ov_h}
    \end{subfigure}  
    \hfill
    \begin{subfigure}{\columnwidth}
        \includegraphics[width=\columnwidth]{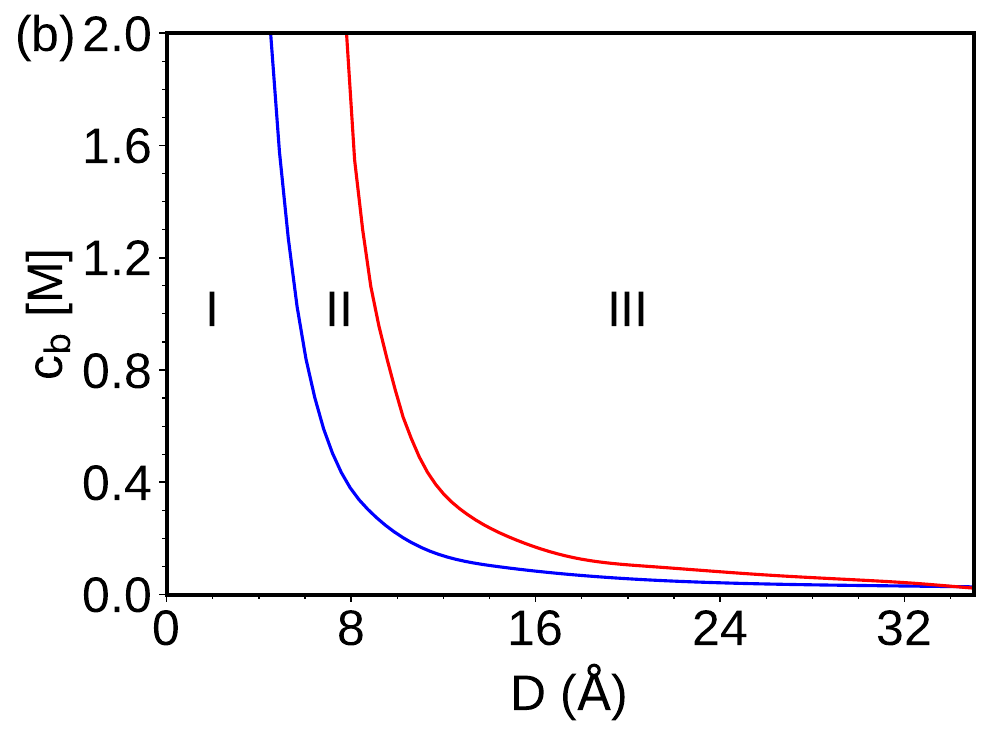}
        \caption{}
        \label{fig:phase_d}
    \end{subfigure} 
\caption{Relationship between opposite-charge repulsion and overcharging. a) Apparent surface charge density of Plate 1, $\sigma_\mathrm{app}(z) = \sigma_1 + \int_0^z \rho(z^*)dz^*$ as a function of the position $z$ for different separation distance $D$ at c$_\mathrm{b} = 0.4 $ M. $\rho(z)$ is the local charge density at $z$. b) Concentration-separation distance phase diagram showing different regions of repulsion, attraction and overcharging. I: no overcharging and no repulsion, II: repulsion without overcharging, III: repulsion with overcharging. $\sigma_1$ = -0.3204 C/m$^2$, $\sigma_2$ = 0.1602 C/m$^2$, $q_+$ = 2, $q_-$ = 1, and $a_\mathrm{+,s} = 1.6$ \AA\ for both the plots.}
\label{fig:ov_ocr}
\end{figure*}

Although previous studies indicated that opposite-charge repulsion is related to overcharging of the surface, the origin and nature of the relationship between
these two phenomena remains unclear to our knowledge. Figure \ref{fig:ov_h} plots the apparent surface charge density $\sigma_\mathrm{app}$ of Plate 1 for divalent salt at $c_\mathrm{b} = 0.4$ M. The divalent cations play the role of the counterion for the negatively charged Plate 1 and will strongly accumulate near the plate surface. $\sigma_\mathrm{app}$ is defined as $\sigma_\mathrm{app}(z) = \sigma_1 + \int_0^z \rho(z^*)dz^*$ with $\rho(z)$ the local net charge density at distance $z$ from plate 1 ($z \in [0,D]$). Thus, $\sigma_\mathrm{app} = \sigma_1$ at $z=0$, whereas $\sigma_\mathrm{app} = -\sigma_2$ at $z=D$. The sign of $\sigma_\mathrm{app}$ signifies the occurrence of overcharging: if $\sigma_\mathrm{app} > 0$ at any position $z$, the electrical double layer is considered overcharged. Based on this definition, profiles for $D = 7$ and $10$ $\text{\AA}$ shown in Figure \ref{fig:ov_h} correspond to normal double layers without overcharging, whereas those for $D = 20$ and $26$ $\text{\AA}$ represent overcharged double layers. On the other hand, the pressure profile for $c_\mathrm{b} = 0.4$ M in Figure \ref{fig:force_conc} shows that the force is attractive at $D = 7\ \text{\AA}$ and turns to repulsive for $D = 10, 20$ and $26$ $\text{\AA}$. The comparison between the double layer structure and force clearly indicates that, at $D = 10\ \text{\AA}$, opposite-charge repulsion occurs in the absence of overcharging. Therefore, the occurrence of opposite-charge repulsion is not necessarily accompanied by overcharging of the double layer.

To fully understand the relationship between opposite-charge repulsion and overcharging, a complete $c_\mathrm{b}$ vs $D$ phase diagram for divalent salt is presented in Figure \ref{fig:phase_d}. The behavior of the double layer structure and force can be divided into three different regions. In Region I, which mostly occupies low separation distances, a normal double layer with attractive force is observed. Both opposite-charge repulsion and overcharging is absent. Region II exist at moderate separation distances, where opposite-charge repulsion is observed without any overcharging. In Region III at large separation distances, both overcharging and opposite-charge repulsion are obtained. The phase diagram clearly shows the existence of the two opposite-charge repulsion regions, Region II and Region III, where the double layer near Plate 1 can either be normal or overcharged. This elucidates the absence of the causal relationship between overcharging and opposite-charge repulsion, in agreement with the simulation results of Trulsson et al.\cite{Trulsson2006RepulsionElectrolytes}.\par 

The enhanced strength of ion-ion correlations in multivalent salt solutions increases the number of both counterions and coions between the two plates, which could lead to both entropy-induced opposite-charge repulsion as well as overcharging. As the system moves from Region III to Region II, the positively charged Plate 2 gets closer to the negatively charged Plate 1, repelling positive multivalent counterions away from the slit. This leads to the disappearance of overcharging. However, the amount of ions in the slit is still large enough to produce a large osmotic pressure and thus a net repulsive force. Finally in Region I, the separation between the plates is too small such that the number of ions in the slit is drastically reduced and the screened electrostatic attraction between the two surfaces dominates.   \par 

\begin{figure}
\includegraphics[width=\columnwidth]{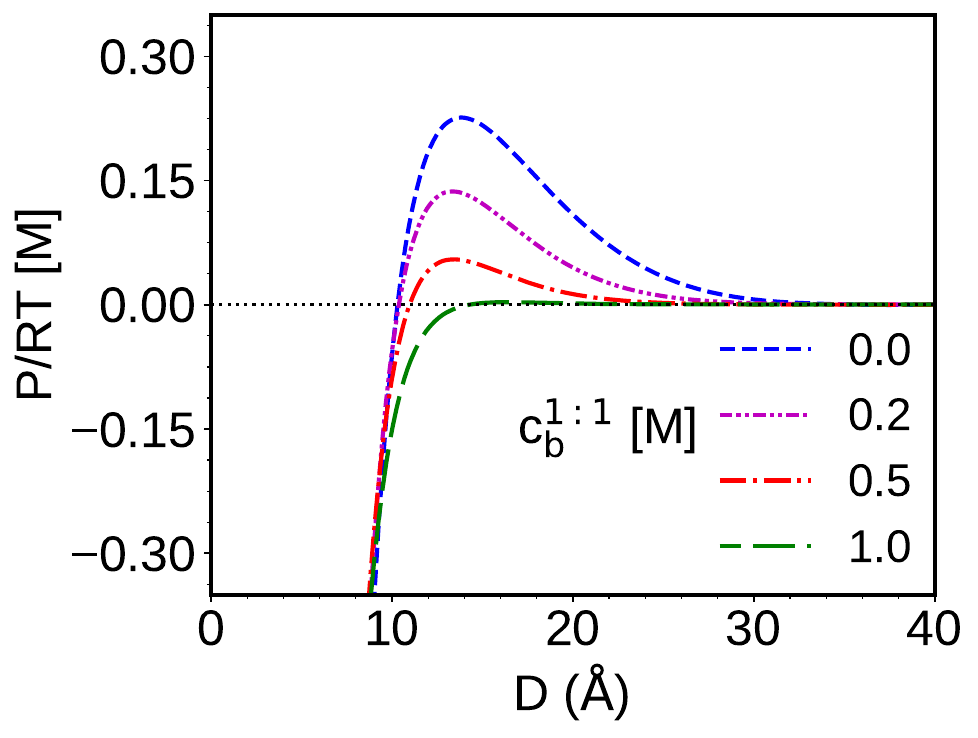}
  \caption{The effect of adding monovalent salt to a solution with a fixed divalent salt concentration on opposite charge repulsion. The the various dashed and dashed-dot lines represent pressures after $c_\mathrm{b}^\mathrm{1:1}$ amount of monovalent salt is added. $c_\mathrm{b}^\mathrm{2:1} = 0.2$ M, $\sigma_1$ = -0.3204 C/m$^2$, $\sigma_2$ = 0.1602 C/m$^2$, and $a_{\pm,s} = 2.0$ \AA.}
  \label{fig:force_mix}
\end{figure}

Finally, we examine the effect of adding monovalent salt on the opposite-charge repulsion, since multivalent and monovalent salt mixtures widely exist in many biological and geological systems. Figure \ref{fig:force_mix} shows that the opposite-charge repulsion is gradually diminished and finally disappears as more and more monovalent salt is added. The force turns to almost completely attractive at $c_\mathrm{b}^\mathrm{1:1} = 1.0$ M. The addition of monovalent salt to multivalent salt solution increases the bulk osmotic pressure. It also reduces the ion-ion correlations due to the lower valency of monovalent ions, leading to the reduction of ion accumulation in the slit. Combining these two effects, the entropic component of the pressure is largely reduced and hence the opposite-charge repulsion.     

\section{Conclusions}

In this work, we apply the modified Gaussian Renormalized Fluctuation theory to study the phenomenon of opposite-charge repulsion and its relationship with overcharging for two planar surfaces immersed in a multivalent salt solution. Both opposite-charge repulsion and overcharging are outcomes of ion-ion correlations. Compared to existing work, our theory accurately accounts for the
effect of the spatially varying ion-ion correlations on the double layer structure and force in a self-consistent
manner. We predict the transition of the force from pure attraction to opposite-charge repulsion upon the addition of multivalent salt, in quantitative agreement with reported simulation results
for all divalent, trivalent, and tetravalent salts. The opposite-charge repulsion is found to be long-ranged with an effective length scale of a few nanometers, which is quite different from the short-range like-charge attraction which acts at a length scale of a few angstroms. The low computational cost of our approach enables us to construct a full concentration-separation distance phase diagram. We examine the unexplored region of medium and high salt concentrations, where we show that the strength of opposite-charge repulsion increases monotonically with the multivalent salt concentration. This indicates different salt concentration dependence of opposite-charge repulsion compared with other ion correlation-driven phenomena, such as overcharging and like-charge attraction, which exhibit non-monotonic behavior. Furthermore, we also predict that the addition of monovalent salt to multivalent salt solutions reduces the strength of the opposite-charge repulsion.\par

We elucidate the electrostatic origin of the opposite-charge repulsion. The enhanced ion-ion correlations in the presence of multivalent ions attract more ions
to the double layer, leading to higher osmotic pressure and stronger screening of the electrostatic attraction. These two effects thus result in an overall repulsive force between the two
oppositely charged surfaces. The same ion-ion correlations also lead to overaccumulation of
counterions in the double layer thus overcharging the surface. However, we demonstrate that there is no causal relationship between
the two phenomena of opposite-charge repulsion and overcharging. In the regime of intermediate separation distance, opposite-charge repulsion can occur in normal double layers without overcharging. This work alongside our previous study on the charge inversion and like-charge attraction, highlights the essential role of ion-ion correlations in determining the structure and properties of electrical double layer, especially in the presence of multivalent ions. Our modified Gaussian Renormalized Fluctuation theory provides a self-consistent quantification of the ion-ion correlations, which paves the way towards the full understanding of the electrostatic interactions in a variety of physical-chemical, soft matter and biological systems.

\begin{acknowledgements}
Acknowledgment is made to the donors of the American Chemical Society Petroleum Research Fund for partial support of this research. N.R.A would like to thank Dr. Keaton J. Burns of MIT, Dr. Adam R. Lamson and Vicente Gómez Herrera of the Flatiron Institute, New York and Prof. Brato Chakrabarti of Flatiron Institute and ICTS, Bengaluru  for introduction to \textit{Dedalus} and assistance in setting up the problem in \textit{Dedalus} framework. The authors also thank Dr. Chao Duan of UC Berkeley for his insightful comments on the results. This research used the Savio computational cluster resource provided by the Berkeley Research Computing program at the University of California, Berkeley (supported by the UC Berkeley Chancellor, Vice Chancellor for Research, and Chief Information Officer).

\end{acknowledgements}
\section*{Data Availability Statement}

The data that support the findings of this study are available from the corresponding author upon reasonable request.

\bibliography{ocr_jcp}

\begin{thebibliography}{70}%
\makeatletter
\providecommand \@ifxundefined [1]{%
 \@ifx{#1\undefined}
}%
\providecommand \@ifnum [1]{%
 \ifnum #1\expandafter \@firstoftwo
 \else \expandafter \@secondoftwo
 \fi
}%
\providecommand \@ifx [1]{%
 \ifx #1\expandafter \@firstoftwo
 \else \expandafter \@secondoftwo
 \fi
}%
\providecommand \natexlab [1]{#1}%
\providecommand \enquote  [1]{``#1''}%
\providecommand \bibnamefont  [1]{#1}%
\providecommand \bibfnamefont [1]{#1}%
\providecommand \citenamefont [1]{#1}%
\providecommand \href@noop [0]{\@secondoftwo}%
\providecommand \href [0]{\begingroup \@sanitize@url \@href}%
\providecommand \@href[1]{\@@startlink{#1}\@@href}%
\providecommand \@@href[1]{\endgroup#1\@@endlink}%
\providecommand \@sanitize@url [0]{\catcode `\\12\catcode `\$12\catcode `\&12\catcode `\#12\catcode `\^12\catcode `\_12\catcode `\%12\relax}%
\providecommand \@@startlink[1]{}%
\providecommand \@@endlink[0]{}%
\providecommand \url  [0]{\begingroup\@sanitize@url \@url }%
\providecommand \@url [1]{\endgroup\@href {#1}{\urlprefix }}%
\providecommand \urlprefix  [0]{URL }%
\providecommand \Eprint [0]{\href }%
\providecommand \doibase [0]{http://dx.doi.org/}%
\providecommand \selectlanguage [0]{\@gobble}%
\providecommand \bibinfo  [0]{\@secondoftwo}%
\providecommand \bibfield  [0]{\@secondoftwo}%
\providecommand \translation [1]{[#1]}%
\providecommand \BibitemOpen [0]{}%
\providecommand \bibitemStop [0]{}%
\providecommand \bibitemNoStop [0]{.\EOS\space}%
\providecommand \EOS [0]{\spacefactor3000\relax}%
\providecommand \BibitemShut  [1]{\csname bibitem#1\endcsname}%
\let\auto@bib@innerbib\@empty
\bibitem [{\citenamefont {Bukosky}\ and\ \citenamefont {Ristenpart}(2015)}]{Bukosky2015SimultaneousFields}%
  \BibitemOpen
  \bibfield  {author} {\bibinfo {author} {\bibfnamefont {S.~C.}\ \bibnamefont {Bukosky}}\ and\ \bibinfo {author} {\bibfnamefont {W.~D.}\ \bibnamefont {Ristenpart}},\ }\bibfield  {title} {\enquote {\bibinfo {title} {{Simultaneous Aggregation and Height Bifurcation of Colloidal Particles near Electrodes in Oscillatory Electric Fields}},}\ }\href {\doibase 10.1021/ACS.LANGMUIR.5B02432/ASSET/IMAGES/LARGE/LA-2015-02432D{\_}0005.JPEG} {\bibfield  {journal} {\bibinfo  {journal} {Langmuir}\ }\textbf {\bibinfo {volume} {31}},\ \bibinfo {pages} {9742--9747} (\bibinfo {year} {2015})}\BibitemShut {NoStop}%
\bibitem [{\citenamefont {Woehl}\ \emph {et~al.}(2015)\citenamefont {Woehl}, \citenamefont {Chen}, \citenamefont {Heatley}, \citenamefont {Talken}, \citenamefont {Bukosky}, \citenamefont {Dutcher},\ and\ \citenamefont {Ristenpart}}]{Woehl2015BifurcationMinimum}%
  \BibitemOpen
  \bibfield  {author} {\bibinfo {author} {\bibfnamefont {T.~J.}\ \bibnamefont {Woehl}}, \bibinfo {author} {\bibfnamefont {B.~J.}\ \bibnamefont {Chen}}, \bibinfo {author} {\bibfnamefont {K.~L.}\ \bibnamefont {Heatley}}, \bibinfo {author} {\bibfnamefont {N.~H.}\ \bibnamefont {Talken}}, \bibinfo {author} {\bibfnamefont {S.~C.}\ \bibnamefont {Bukosky}}, \bibinfo {author} {\bibfnamefont {C.~S.}\ \bibnamefont {Dutcher}}, \ and\ \bibinfo {author} {\bibfnamefont {W.~D.}\ \bibnamefont {Ristenpart}},\ }\bibfield  {title} {\enquote {\bibinfo {title} {{Bifurcation in the steady-state height of colloidal particles near an electrode in oscillatory electric fields: Evidence for a tertiary potential minimum}},}\ }\href {\doibase 10.1103/PHYSREVX.5.011023/FIGURES/6/MEDIUM} {\bibfield  {journal} {\bibinfo  {journal} {Phys. Rev. X}\ }\textbf {\bibinfo {volume} {5}},\ \bibinfo {pages} {011023} (\bibinfo {year} {2015})}\BibitemShut {NoStop}%
\bibitem [{\citenamefont {Ajdari}(2000)}]{Ajdari2000PumpingArrays}%
  \BibitemOpen
  \bibfield  {author} {\bibinfo {author} {\bibfnamefont {A.}~\bibnamefont {Ajdari}},\ }\bibfield  {title} {\enquote {\bibinfo {title} {{Pumping liquids using asymmetric electrode arrays}},}\ }\href {\doibase 10.1103/PhysRevE.61.R45} {\bibfield  {journal} {\bibinfo  {journal} {Phys. Rev. E}\ }\textbf {\bibinfo {volume} {61}},\ \bibinfo {pages} {R45--R48} (\bibinfo {year} {2000})}\BibitemShut {NoStop}%
\bibitem [{\citenamefont {Itskovich}, \citenamefont {Kornyshev},\ and\ \citenamefont {Vorotyntsev}(1977)}]{Itskovich1977ElectricCharacteristic}%
  \BibitemOpen
  \bibfield  {author} {\bibinfo {author} {\bibfnamefont {E.~M.}\ \bibnamefont {Itskovich}}, \bibinfo {author} {\bibfnamefont {A.~A.}\ \bibnamefont {Kornyshev}}, \ and\ \bibinfo {author} {\bibfnamefont {M.~A.}\ \bibnamefont {Vorotyntsev}},\ }\bibfield  {title} {\enquote {\bibinfo {title} {{Electric current across the metal–solid electrolyte interface I. Direct current, current–voltage characteristic}},}\ }\href {\doibase 10.1002/pssa.2210390126} {\bibfield  {journal} {\bibinfo  {journal} {Phys. Status Solidi A}\ }\textbf {\bibinfo {volume} {39}},\ \bibinfo {pages} {229--238} (\bibinfo {year} {1977})}\BibitemShut {NoStop}%
\bibitem [{\citenamefont {Israelachvili}(2011)}]{israelachvili2011intermolecular}%
  \BibitemOpen
  \bibfield  {author} {\bibinfo {author} {\bibfnamefont {J.~N.}\ \bibnamefont {Israelachvili}},\ }\href@noop {} {\emph {\bibinfo {title} {Intermolecular and surface forces}}}\ (\bibinfo  {publisher} {Academic press: New York},\ \bibinfo {year} {2011})\BibitemShut {NoStop}%
\bibitem [{\citenamefont {Larsen}\ and\ \citenamefont {Grier}(1997)}]{Larsen1997Like-chargeCrystallites}%
  \BibitemOpen
  \bibfield  {author} {\bibinfo {author} {\bibfnamefont {A.~E.}\ \bibnamefont {Larsen}}\ and\ \bibinfo {author} {\bibfnamefont {D.~G.}\ \bibnamefont {Grier}},\ }\bibfield  {title} {\enquote {\bibinfo {title} {{Like-charge attractions in metastable colloidal crystallites}},}\ }\href {\doibase 10.1038/385230a0} {\bibfield  {journal} {\bibinfo  {journal} {Nature}\ }\textbf {\bibinfo {volume} {385}},\ \bibinfo {pages} {230--233} (\bibinfo {year} {1997})}\BibitemShut {NoStop}%
\bibitem [{\citenamefont {Dinsmore}, \citenamefont {Crocker},\ and\ \citenamefont {Yodh}(1998)}]{Dinsmore1998Self-assemblyCrystals}%
  \BibitemOpen
  \bibfield  {author} {\bibinfo {author} {\bibfnamefont {A.~D.}\ \bibnamefont {Dinsmore}}, \bibinfo {author} {\bibfnamefont {J.~C.}\ \bibnamefont {Crocker}}, \ and\ \bibinfo {author} {\bibfnamefont {A.~G.}\ \bibnamefont {Yodh}},\ }\bibfield  {title} {\enquote {\bibinfo {title} {{Self-assembly of colloidal crystals}},}\ }\href {\doibase 10.1016/S1359-0294(98)80035-6} {\bibfield  {journal} {\bibinfo  {journal} {Curr. Opin. Colloid Interface Sci.}\ }\textbf {\bibinfo {volume} {3}},\ \bibinfo {pages} {5--11} (\bibinfo {year} {1998})}\BibitemShut {NoStop}%
\bibitem [{\citenamefont {Caccamo}, \citenamefont {Pellicane},\ and\ \citenamefont {Costa}(2000)}]{Caccamo_2000}%
  \BibitemOpen
  \bibfield  {author} {\bibinfo {author} {\bibfnamefont {C.}~\bibnamefont {Caccamo}}, \bibinfo {author} {\bibfnamefont {G.}~\bibnamefont {Pellicane}}, \ and\ \bibinfo {author} {\bibfnamefont {D.}~\bibnamefont {Costa}},\ }\bibfield  {title} {\enquote {\bibinfo {title} {Phase transitions in hard-core yukawa fluids: toward a theory of phase stability in protein solutions},}\ }\href {\doibase 10.1088/0953-8984/12/8a/360} {\bibfield  {journal} {\bibinfo  {journal} {J. Phys. Condens. Matter}\ }\textbf {\bibinfo {volume} {12}},\ \bibinfo {pages} {A437--A442} (\bibinfo {year} {2000})}\BibitemShut {NoStop}%
\bibitem [{\citenamefont {Butler}\ \emph {et~al.}(2003)\citenamefont {Butler}, \citenamefont {Angelini}, \citenamefont {Tang},\ and\ \citenamefont {Wong}}]{Butler2003IonAttraction}%
  \BibitemOpen
  \bibfield  {author} {\bibinfo {author} {\bibfnamefont {J.~C.}\ \bibnamefont {Butler}}, \bibinfo {author} {\bibfnamefont {T.}~\bibnamefont {Angelini}}, \bibinfo {author} {\bibfnamefont {J.~X.}\ \bibnamefont {Tang}}, \ and\ \bibinfo {author} {\bibfnamefont {G.~C.}\ \bibnamefont {Wong}},\ }\bibfield  {title} {\enquote {\bibinfo {title} {{Ion Multivalence and Like-Charge Polyelectrolyte Attraction}},}\ }\href {\doibase 10.1103/PhysRevLett.91.028301} {\bibfield  {journal} {\bibinfo  {journal} {Phys. Rev. Lett.}\ }\textbf {\bibinfo {volume} {91}} (\bibinfo {year} {2003}),\ 10.1103/PhysRevLett.91.028301}\BibitemShut {NoStop}%
\bibitem [{\citenamefont {Zhang}\ \emph {et~al.}(2008)\citenamefont {Zhang}, \citenamefont {Skoda}, \citenamefont {Jacobs}, \citenamefont {Zorn}, \citenamefont {Martin}, \citenamefont {Martin}, \citenamefont {Clark}, \citenamefont {Weggler}, \citenamefont {Hildebrandt}, \citenamefont {Kohlbacher},\ and\ \citenamefont {Schreiber}}]{Zhang2008ReentrantCounterions}%
  \BibitemOpen
  \bibfield  {author} {\bibinfo {author} {\bibfnamefont {F.}~\bibnamefont {Zhang}}, \bibinfo {author} {\bibfnamefont {M.~W.}\ \bibnamefont {Skoda}}, \bibinfo {author} {\bibfnamefont {R.~M.}\ \bibnamefont {Jacobs}}, \bibinfo {author} {\bibfnamefont {S.}~\bibnamefont {Zorn}}, \bibinfo {author} {\bibfnamefont {R.~A.}\ \bibnamefont {Martin}}, \bibinfo {author} {\bibfnamefont {C.~M.}\ \bibnamefont {Martin}}, \bibinfo {author} {\bibfnamefont {G.~F.}\ \bibnamefont {Clark}}, \bibinfo {author} {\bibfnamefont {S.}~\bibnamefont {Weggler}}, \bibinfo {author} {\bibfnamefont {A.}~\bibnamefont {Hildebrandt}}, \bibinfo {author} {\bibfnamefont {O.}~\bibnamefont {Kohlbacher}}, \ and\ \bibinfo {author} {\bibfnamefont {F.}~\bibnamefont {Schreiber}},\ }\bibfield  {title} {\enquote {\bibinfo {title} {{Reentrant condensation of proteins in solution induced by multivalent counterions}},}\ }\href {\doibase 10.1103/PhysRevLett.101.148101} {\bibfield  {journal} {\bibinfo  {journal} {Phys. Rev. Lett.}\ }\textbf {\bibinfo {volume} {101}}
  (\bibinfo {year} {2008}),\ 10.1103/PhysRevLett.101.148101}\BibitemShut {NoStop}%
\bibitem [{\citenamefont {Nassar}\ \emph {et~al.}(2021)\citenamefont {Nassar}, \citenamefont {Dignon}, \citenamefont {Razban},\ and\ \citenamefont {Dill}}]{Nassar2021TheTheory}%
  \BibitemOpen
  \bibfield  {author} {\bibinfo {author} {\bibfnamefont {R.}~\bibnamefont {Nassar}}, \bibinfo {author} {\bibfnamefont {G.~L.}\ \bibnamefont {Dignon}}, \bibinfo {author} {\bibfnamefont {R.~M.}\ \bibnamefont {Razban}}, \ and\ \bibinfo {author} {\bibfnamefont {K.~A.}\ \bibnamefont {Dill}},\ }\bibfield  {title} {\enquote {\bibinfo {title} {{The Protein Folding Problem: The Role of Theory}},}\ }\href {\doibase 10.1016/J.JMB.2021.167126} {\bibfield  {journal} {\bibinfo  {journal} {J. Mol. Biol.}\ }\textbf {\bibinfo {volume} {433}} (\bibinfo {year} {2021}),\ 10.1016/J.JMB.2021.167126}\BibitemShut {NoStop}%
\bibitem [{\citenamefont {Cevc}(1990)}]{Cevc1990MembraneElectrostatics}%
  \BibitemOpen
  \bibfield  {author} {\bibinfo {author} {\bibfnamefont {G.}~\bibnamefont {Cevc}},\ }\bibfield  {title} {\enquote {\bibinfo {title} {Membrane electrostatics},}\ }\href {\doibase https://doi.org/10.1016/0304-4157(90)90015-5} {\bibfield  {journal} {\bibinfo  {journal} {Biochim. Biophys. Acta - Biomembr.}\ }\textbf {\bibinfo {volume} {1031}},\ \bibinfo {pages} {311--382} (\bibinfo {year} {1990})}\BibitemShut {NoStop}%
\bibitem [{\citenamefont {Kozlov}\ \emph {et~al.}(2014)\citenamefont {Kozlov}, \citenamefont {Campelo}, \citenamefont {Liska}, \citenamefont {Chernomordik}, \citenamefont {Marrink},\ and\ \citenamefont {McMahon}}]{Kozlov2014MechanismsMembranes}%
  \BibitemOpen
  \bibfield  {author} {\bibinfo {author} {\bibfnamefont {M.~M.}\ \bibnamefont {Kozlov}}, \bibinfo {author} {\bibfnamefont {F.}~\bibnamefont {Campelo}}, \bibinfo {author} {\bibfnamefont {N.}~\bibnamefont {Liska}}, \bibinfo {author} {\bibfnamefont {L.~V.}\ \bibnamefont {Chernomordik}}, \bibinfo {author} {\bibfnamefont {S.~J.}\ \bibnamefont {Marrink}}, \ and\ \bibinfo {author} {\bibfnamefont {H.~T.}\ \bibnamefont {McMahon}},\ }\bibfield  {title} {\enquote {\bibinfo {title} {{Mechanisms shaping cell membranes}},}\ }\href {\doibase 10.1016/J.CEB.2014.03.006} {\bibfield  {journal} {\bibinfo  {journal} {Curr. Opin. Cell Biol.}\ }\textbf {\bibinfo {volume} {29}},\ \bibinfo {pages} {53--60} (\bibinfo {year} {2014})}\BibitemShut {NoStop}%
\bibitem [{\citenamefont {K{\'{e}}kicheff}\ \emph {et~al.}(1993)\citenamefont {K{\'{e}}kicheff}, \citenamefont {Mar\u{c}elja}, \citenamefont {Senden},\ and\ \citenamefont {Shubin}}]{Kekicheff1993ChargeElectrolyte}%
  \BibitemOpen
  \bibfield  {author} {\bibinfo {author} {\bibfnamefont {P.}~\bibnamefont {K{\'{e}}kicheff}}, \bibinfo {author} {\bibfnamefont {S.}~\bibnamefont {Mar\u{c}elja}}, \bibinfo {author} {\bibfnamefont {T.~J.}\ \bibnamefont {Senden}}, \ and\ \bibinfo {author} {\bibfnamefont {V.~E.}\ \bibnamefont {Shubin}},\ }\bibfield  {title} {\enquote {\bibinfo {title} {{Charge reversal seen in electrical double layer interaction of surfaces immersed in 2:1 calcium electrolyte}},}\ }\href {\doibase 10.1063/1.465906} {\bibfield  {journal} {\bibinfo  {journal} {J. Chem. Phys.}\ }\textbf {\bibinfo {volume} {99}},\ \bibinfo {pages} {6098} (\bibinfo {year} {1993})}\BibitemShut {NoStop}%
\bibitem [{\citenamefont {Zohar}, \citenamefont {Leizerson},\ and\ \citenamefont {Sivan}(2006)}]{Zohar2006ShortSurfaces}%
  \BibitemOpen
  \bibfield  {author} {\bibinfo {author} {\bibfnamefont {O.}~\bibnamefont {Zohar}}, \bibinfo {author} {\bibfnamefont {I.}~\bibnamefont {Leizerson}}, \ and\ \bibinfo {author} {\bibfnamefont {U.}~\bibnamefont {Sivan}},\ }\bibfield  {title} {\enquote {\bibinfo {title} {{Short range attraction between two similarly charged silica surfaces}},}\ }\href {\doibase 10.1103/PhysRevLett.96.177802} {\bibfield  {journal} {\bibinfo  {journal} {Phys. Rev. Lett.}\ }\textbf {\bibinfo {volume} {96}} (\bibinfo {year} {2006}),\ 10.1103/PhysRevLett.96.177802}\BibitemShut {NoStop}%
\bibitem [{\citenamefont {Kumar}\ \emph {et~al.}(2017)\citenamefont {Kumar}, \citenamefont {Yadav}, \citenamefont {Abbas}, \citenamefont {Aswal},\ and\ \citenamefont {Kohlbrecher}}]{Kumar2017InteractionsIons}%
  \BibitemOpen
  \bibfield  {author} {\bibinfo {author} {\bibfnamefont {S.}~\bibnamefont {Kumar}}, \bibinfo {author} {\bibfnamefont {I.}~\bibnamefont {Yadav}}, \bibinfo {author} {\bibfnamefont {S.}~\bibnamefont {Abbas}}, \bibinfo {author} {\bibfnamefont {V.~K.}\ \bibnamefont {Aswal}}, \ and\ \bibinfo {author} {\bibfnamefont {J.}~\bibnamefont {Kohlbrecher}},\ }\bibfield  {title} {\enquote {\bibinfo {title} {{Interactions in reentrant phase behavior of a charged nanoparticle solution by multivalent ions}},}\ }\href {\doibase 10.1103/PhysRevE.96.060602} {\bibfield  {journal} {\bibinfo  {journal} {Phys. Rev. E}\ }\textbf {\bibinfo {volume} {96}},\ \bibinfo {pages} {60602} (\bibinfo {year} {2017})}\BibitemShut {NoStop}%
\bibitem [{\citenamefont {Linse}\ and\ \citenamefont {Lobaskin}(1999)}]{linselca1999}%
  \BibitemOpen
  \bibfield  {author} {\bibinfo {author} {\bibfnamefont {P.}~\bibnamefont {Linse}}\ and\ \bibinfo {author} {\bibfnamefont {V.}~\bibnamefont {Lobaskin}},\ }\bibfield  {title} {\enquote {\bibinfo {title} {Electrostatic attraction and phase separation in solutions of like-charged colloidal particles},}\ }\href {\doibase 10.1103/PhysRevLett.83.4208} {\bibfield  {journal} {\bibinfo  {journal} {Phys. Rev. Lett.}\ }\textbf {\bibinfo {volume} {83}},\ \bibinfo {pages} {4208--4211} (\bibinfo {year} {1999})}\BibitemShut {NoStop}%
\bibitem [{\citenamefont {Linse}\ and\ \citenamefont {Lobaskin}(2000)}]{Linse2000}%
  \BibitemOpen
  \bibfield  {author} {\bibinfo {author} {\bibfnamefont {P.}~\bibnamefont {Linse}}\ and\ \bibinfo {author} {\bibfnamefont {V.}~\bibnamefont {Lobaskin}},\ }\bibfield  {title} {\enquote {\bibinfo {title} {{Electrostatic attraction and phase separation in solutions of like-charged colloidal particles}},}\ }\href {\doibase 10.1063/1.480943} {\bibfield  {journal} {\bibinfo  {journal} {J. Chem. Phys}\ }\textbf {\bibinfo {volume} {112}},\ \bibinfo {pages} {3917--3927} (\bibinfo {year} {2000})}\BibitemShut {NoStop}%
\bibitem [{\citenamefont {Wu}\ \emph {et~al.}(1999)\citenamefont {Wu}, \citenamefont {Bratko}, \citenamefont {Blanch},\ and\ \citenamefont {Prausnitz}}]{Wu1999MonteSalts}%
  \BibitemOpen
  \bibfield  {author} {\bibinfo {author} {\bibfnamefont {J.~Z.}\ \bibnamefont {Wu}}, \bibinfo {author} {\bibfnamefont {D.}~\bibnamefont {Bratko}}, \bibinfo {author} {\bibfnamefont {H.~W.}\ \bibnamefont {Blanch}}, \ and\ \bibinfo {author} {\bibfnamefont {J.~M.}\ \bibnamefont {Prausnitz}},\ }\bibfield  {title} {\enquote {\bibinfo {title} {{Monte Carlo simulation for the potential of mean force between ionic colloids in solutions of asymmetric salts}},}\ }\href {\doibase 10.1063/1.480000} {\bibfield  {journal} {\bibinfo  {journal} {J. Chem. Phys}\ }\textbf {\bibinfo {volume} {111}},\ \bibinfo {pages} {7084--7094} (\bibinfo {year} {1999})}\BibitemShut {NoStop}%
\bibitem [{\citenamefont {Angelescu}\ and\ \citenamefont {Linse}(2003)}]{Angelescu2003MonteAdded}%
  \BibitemOpen
  \bibfield  {author} {\bibinfo {author} {\bibfnamefont {D.~G.}\ \bibnamefont {Angelescu}}\ and\ \bibinfo {author} {\bibfnamefont {P.}~\bibnamefont {Linse}},\ }\bibfield  {title} {\enquote {\bibinfo {title} {Monte carlo simulation of the mean force between two like-charged macroions with simple 1: 3 salt added},}\ }\href {\doibase 10.1021/la030153a} {\bibfield  {journal} {\bibinfo  {journal} {Langmuir}\ }\textbf {\bibinfo {volume} {19}},\ \bibinfo {pages} {9661--9668} (\bibinfo {year} {2003})}\BibitemShut {NoStop}%
\bibitem [{\citenamefont {Zhang}\ \emph {et~al.}(2016)\citenamefont {Zhang}, \citenamefont {Zhang}, \citenamefont {Shi}, \citenamefont {Zhu},\ and\ \citenamefont {Tan}}]{Zhang2016PotentialEffect}%
  \BibitemOpen
  \bibfield  {author} {\bibinfo {author} {\bibfnamefont {X.}~\bibnamefont {Zhang}}, \bibinfo {author} {\bibfnamefont {J.~S.}\ \bibnamefont {Zhang}}, \bibinfo {author} {\bibfnamefont {Y.~Z.}\ \bibnamefont {Shi}}, \bibinfo {author} {\bibfnamefont {X.~L.}\ \bibnamefont {Zhu}}, \ and\ \bibinfo {author} {\bibfnamefont {Z.~J.}\ \bibnamefont {Tan}},\ }\bibfield  {title} {\enquote {\bibinfo {title} {{Potential of mean force between like-charged nanoparticles: Many-body effect}},}\ }\href {\doibase 10.1038/srep23434} {\bibfield  {journal} {\bibinfo  {journal} {Sci. Rep}\ }\textbf {\bibinfo {volume} {6}},\ \bibinfo {pages} {1--12} (\bibinfo {year} {2016})}\BibitemShut {NoStop}%
\bibitem [{\citenamefont {Besteman}\ \emph {et~al.}(2004)\citenamefont {Besteman}, \citenamefont {Zevenbergen}, \citenamefont {Heering},\ and\ \citenamefont {Lemay}}]{Besteman2004DirectPhenomenon}%
  \BibitemOpen
  \bibfield  {author} {\bibinfo {author} {\bibfnamefont {K.}~\bibnamefont {Besteman}}, \bibinfo {author} {\bibfnamefont {M.~A.}\ \bibnamefont {Zevenbergen}}, \bibinfo {author} {\bibfnamefont {H.~A.}\ \bibnamefont {Heering}}, \ and\ \bibinfo {author} {\bibfnamefont {S.~G.}\ \bibnamefont {Lemay}},\ }\bibfield  {title} {\enquote {\bibinfo {title} {{Direct observation of charge inversion by multivalent ions as a universal electrostatic phenomenon}},}\ }\href {\doibase 10.1103/PhysRevLett.93.170802} {\bibfield  {journal} {\bibinfo  {journal} {Phys. Rev. Lett.}\ }\textbf {\bibinfo {volume} {93}},\ \bibinfo {pages} {170802} (\bibinfo {year} {2004})}\BibitemShut {NoStop}%
\bibitem [{\citenamefont {Besteman}, \citenamefont {Zevenbergen},\ and\ \citenamefont {Lemay}(2005)}]{Besteman2005ChargeDensity}%
  \BibitemOpen
  \bibfield  {author} {\bibinfo {author} {\bibfnamefont {K.}~\bibnamefont {Besteman}}, \bibinfo {author} {\bibfnamefont {M.~A.~G.}\ \bibnamefont {Zevenbergen}}, \ and\ \bibinfo {author} {\bibfnamefont {S.~G.}\ \bibnamefont {Lemay}},\ }\bibfield  {title} {\enquote {\bibinfo {title} {{Charge inversion by multivalent ions: Dependence on dielectric constant and surface-charge density}},}\ }\href {\doibase 10.1103/PhysRevE.72.061501} {\bibfield  {journal} {\bibinfo  {journal} {Phys. Rev. E}\ }\textbf {\bibinfo {volume} {72}} (\bibinfo {year} {2005}),\ 10.1103/PhysRevE.72.061501}\BibitemShut {NoStop}%
\bibitem [{\citenamefont {Trulsson}\ \emph {et~al.}(2006)\citenamefont {Trulsson}, \citenamefont {J{\"{o}}nsson}, \citenamefont {{\AA}kesson}, \citenamefont {Forsman},\ and\ \citenamefont {Labbez}}]{Trulsson2006RepulsionElectrolytes}%
  \BibitemOpen
  \bibfield  {author} {\bibinfo {author} {\bibfnamefont {M.}~\bibnamefont {Trulsson}}, \bibinfo {author} {\bibfnamefont {B.}~\bibnamefont {J{\"{o}}nsson}}, \bibinfo {author} {\bibfnamefont {T.}~\bibnamefont {{\AA}kesson}}, \bibinfo {author} {\bibfnamefont {J.}~\bibnamefont {Forsman}}, \ and\ \bibinfo {author} {\bibfnamefont {C.}~\bibnamefont {Labbez}},\ }\bibfield  {title} {\enquote {\bibinfo {title} {{Repulsion between oppositely charged surfaces in multivalent electrolytes}},}\ }\href {\doibase 10.1103/PhysRevLett.97.068302} {\bibfield  {journal} {\bibinfo  {journal} {Phys. Rev. Lett.}\ }\textbf {\bibinfo {volume} {97}} (\bibinfo {year} {2006}),\ 10.1103/PhysRevLett.97.068302}\BibitemShut {NoStop}%
\bibitem [{\citenamefont {Trulsson}\ \emph {et~al.}(2007)\citenamefont {Trulsson}, \citenamefont {J{\"{o}}nsson}, \citenamefont {{\AA}kesson}, \citenamefont {Forsman},\ and\ \citenamefont {Labbez}}]{Trulsson2007RepulsionParticles}%
  \BibitemOpen
  \bibfield  {author} {\bibinfo {author} {\bibfnamefont {M.}~\bibnamefont {Trulsson}}, \bibinfo {author} {\bibfnamefont {B.}~\bibnamefont {J{\"{o}}nsson}}, \bibinfo {author} {\bibfnamefont {T.}~\bibnamefont {{\AA}kesson}}, \bibinfo {author} {\bibfnamefont {J.}~\bibnamefont {Forsman}}, \ and\ \bibinfo {author} {\bibfnamefont {C.}~\bibnamefont {Labbez}},\ }\bibfield  {title} {\enquote {\bibinfo {title} {{Repulsion between oppositely charged macromolecules or particles}},}\ }\href {\doibase 10.1021/LA701222B/ASSET/IMAGES/LARGE/LA701222BF00011.JPEG} {\bibfield  {journal} {\bibinfo  {journal} {Langmuir}\ }\textbf {\bibinfo {volume} {23}},\ \bibinfo {pages} {11562--11569} (\bibinfo {year} {2007})}\BibitemShut {NoStop}%
\bibitem [{\citenamefont {Popa}\ \emph {et~al.}(2010)\citenamefont {Popa}, \citenamefont {Sinha}, \citenamefont {Finessi}, \citenamefont {Maroni}, \citenamefont {Papastavrou},\ and\ \citenamefont {Borkovec}}]{Popa2010ImportanceSurfaces}%
  \BibitemOpen
  \bibfield  {author} {\bibinfo {author} {\bibfnamefont {I.}~\bibnamefont {Popa}}, \bibinfo {author} {\bibfnamefont {P.}~\bibnamefont {Sinha}}, \bibinfo {author} {\bibfnamefont {M.}~\bibnamefont {Finessi}}, \bibinfo {author} {\bibfnamefont {P.}~\bibnamefont {Maroni}}, \bibinfo {author} {\bibfnamefont {G.}~\bibnamefont {Papastavrou}}, \ and\ \bibinfo {author} {\bibfnamefont {M.}~\bibnamefont {Borkovec}},\ }\bibfield  {title} {\enquote {\bibinfo {title} {{Importance of charge regulation in attractive double-layer forces between dissimilar surfaces}},}\ }\href {\doibase 10.1103/PhysRevLett.104.228301} {\bibfield  {journal} {\bibinfo  {journal} {Phys. Rev. Lett.}\ }\textbf {\bibinfo {volume} {104}},\ \bibinfo {pages} {228301--228302} (\bibinfo {year} {2010})}\BibitemShut {NoStop}%
\bibitem [{\citenamefont {Montes Ruiz-Cabello}\ \emph {et~al.}(2014)\citenamefont {Montes Ruiz-Cabello}, \citenamefont {Trefalt}, \citenamefont {Maroni},\ and\ \citenamefont {Borkovec}}]{MontesRuiz-Cabello2014AccurateTheory}%
  \BibitemOpen
  \bibfield  {author} {\bibinfo {author} {\bibfnamefont {F.~J.}\ \bibnamefont {Montes Ruiz-Cabello}}, \bibinfo {author} {\bibfnamefont {G.}~\bibnamefont {Trefalt}}, \bibinfo {author} {\bibfnamefont {P.}~\bibnamefont {Maroni}}, \ and\ \bibinfo {author} {\bibfnamefont {M.}~\bibnamefont {Borkovec}},\ }\bibfield  {title} {\enquote {\bibinfo {title} {Accurate predictions of forces in the presence of multivalent ions by poisson–boltzmann theory},}\ }\href {\doibase 10.1021/la500612a} {\bibfield  {journal} {\bibinfo  {journal} {Langmuir}\ }\textbf {\bibinfo {volume} {30}},\ \bibinfo {pages} {4551--4555} (\bibinfo {year} {2014})},\ \Eprint {http://arxiv.org/abs/https://doi.org/10.1021/la500612a} {https://doi.org/10.1021/la500612a} \BibitemShut {NoStop}%
\bibitem [{\citenamefont {Antila}, \citenamefont {Van~Tassel},\ and\ \citenamefont {Sammalkorpi}(2016)}]{AntilaInteractionRods}%
  \BibitemOpen
  \bibfield  {author} {\bibinfo {author} {\bibfnamefont {H.~S.}\ \bibnamefont {Antila}}, \bibinfo {author} {\bibfnamefont {P.~R.}\ \bibnamefont {Van~Tassel}}, \ and\ \bibinfo {author} {\bibfnamefont {M.}~\bibnamefont {Sammalkorpi}},\ }\bibfield  {title} {\enquote {\bibinfo {title} {Interaction modes between asymmetrically and oppositely charged rods},}\ }\href {\doibase 10.1103/PhysRevE.93.022602} {\bibfield  {journal} {\bibinfo  {journal} {Phys. Rev. E}\ }\textbf {\bibinfo {volume} {93}},\ \bibinfo {pages} {022602} (\bibinfo {year} {2016})}\BibitemShut {NoStop}%
\bibitem [{\citenamefont {Antila}, \citenamefont {Van~Tassel},\ and\ \citenamefont {Sammalkorpi}(2017)}]{Antila2017RepulsionConfinement}%
  \BibitemOpen
  \bibfield  {author} {\bibinfo {author} {\bibfnamefont {H.~S.}\ \bibnamefont {Antila}}, \bibinfo {author} {\bibfnamefont {P.~R.}\ \bibnamefont {Van~Tassel}}, \ and\ \bibinfo {author} {\bibfnamefont {M.}~\bibnamefont {Sammalkorpi}},\ }\bibfield  {title} {\enquote {\bibinfo {title} {{Repulsion between oppositely charged rod-shaped macromolecules: Role of overcharging and ionic confinement}},}\ }\href {\doibase 10.1063/1.4993492/1006544} {\bibfield  {journal} {\bibinfo  {journal} {J. Chem. Phys.}\ }\textbf {\bibinfo {volume} {147}} (\bibinfo {year} {2017}),\ 10.1063/1.4993492/1006544}\BibitemShut {NoStop}%
\bibitem [{\citenamefont {Moazzami-Gudarzi}\ \emph {et~al.}(2018)\citenamefont {Moazzami-Gudarzi}, \citenamefont {Pavel~Adam}, \citenamefont {Smith}, \citenamefont {Trefalt}, \citenamefont {Szil{\'{a}}~gyi}, \citenamefont {Maroni},\ and\ \citenamefont {Borkovec}}]{Moazzami-Gudarzi2018CiteThis}%
  \BibitemOpen
  \bibfield  {author} {\bibinfo {author} {\bibfnamefont {M.}~\bibnamefont {Moazzami-Gudarzi}}, \bibinfo {author} {\bibfnamefont {a.}~\bibnamefont {Pavel~Adam}}, \bibinfo {author} {\bibfnamefont {A.~M.}\ \bibnamefont {Smith}}, \bibinfo {author} {\bibfnamefont {G.}~\bibnamefont {Trefalt}}, \bibinfo {author} {\bibfnamefont {I.}~\bibnamefont {Szil{\'{a}}~gyi}}, \bibinfo {author} {\bibfnamefont {P.}~\bibnamefont {Maroni}}, \ and\ \bibinfo {author} {\bibfnamefont {M.}~\bibnamefont {Borkovec}},\ }\bibfield  {title} {\enquote {\bibinfo {title} {{Cite this}},}\ }\href {\doibase 10.1039/c8cp00679b} {\bibfield  {journal} {\bibinfo  {journal} {Phys. Chem. Chem. Phys}\ }\textbf {\bibinfo {volume} {20}},\ \bibinfo {pages} {9436} (\bibinfo {year} {2018})}\BibitemShut {NoStop}%
\bibitem [{\citenamefont {Lin}\ \emph {et~al.}(2019)\citenamefont {Lin}, \citenamefont {Zhang}, \citenamefont {Qiang}, \citenamefont {Zhang},\ and\ \citenamefont {Tan}}]{Lin2019ApparentSolutions}%
  \BibitemOpen
  \bibfield  {author} {\bibinfo {author} {\bibfnamefont {C.}~\bibnamefont {Lin}}, \bibinfo {author} {\bibfnamefont {X.}~\bibnamefont {Zhang}}, \bibinfo {author} {\bibfnamefont {X.}~\bibnamefont {Qiang}}, \bibinfo {author} {\bibfnamefont {J.~S.}\ \bibnamefont {Zhang}}, \ and\ \bibinfo {author} {\bibfnamefont {Z.~J.}\ \bibnamefont {Tan}},\ }\bibfield  {title} {\enquote {\bibinfo {title} {{Apparent repulsion between equally and oppositely charged spherical polyelectrolytes in symmetrical salt solutions}},}\ }\href {\doibase 10.1063/1.5120756} {\bibfield  {journal} {\bibinfo  {journal} {J. Chem. Phys.}\ }\textbf {\bibinfo {volume} {151}},\ \bibinfo {pages} {114902} (\bibinfo {year} {2019})}\BibitemShut {NoStop}%
\bibitem [{\citenamefont {Patey}(1980)}]{Patey1980TheApproximation}%
  \BibitemOpen
  \bibfield  {author} {\bibinfo {author} {\bibfnamefont {G.~N.}\ \bibnamefont {Patey}},\ }\bibfield  {title} {\enquote {\bibinfo {title} {{The interaction of two spherical colloidal particles in electrolyte solution. An application of the hypernetted-chain approximation}},}\ }\href {\doibase 10.1063/1.438997} {\bibfield  {journal} {\bibinfo  {journal} {J. Chem. Phys}\ }\textbf {\bibinfo {volume} {72}},\ \bibinfo {pages} {5763} (\bibinfo {year} {1980})}\BibitemShut {NoStop}%
\bibitem [{\citenamefont {Kjellander}\ \emph {et~al.}(1988)\citenamefont {Kjellander}, \citenamefont {Mar{\v{c}}elja}, \citenamefont {Pashley},\ and\ \citenamefont {Quirk}}]{Kjellander1988Double-LayerSwelling}%
  \BibitemOpen
  \bibfield  {author} {\bibinfo {author} {\bibfnamefont {R.}~\bibnamefont {Kjellander}}, \bibinfo {author} {\bibfnamefont {S.}~\bibnamefont {Mar{\v{c}}elja}}, \bibinfo {author} {\bibfnamefont {R.~M.}\ \bibnamefont {Pashley}}, \ and\ \bibinfo {author} {\bibfnamefont {J.~P.}\ \bibnamefont {Quirk}},\ }\bibfield  {title} {\enquote {\bibinfo {title} {{Double-layer ion correlation forces restrict calcium-clay swelling}},}\ }\href {\doibase 10.1021/j100334a005} {\bibfield  {journal} {\bibinfo  {journal} {J. Phys. Chem.}\ }\textbf {\bibinfo {volume} {92}},\ \bibinfo {pages} {6489--6492} (\bibinfo {year} {1988})}\BibitemShut {NoStop}%
\bibitem [{\citenamefont {Rouzina}\ and\ \citenamefont {Bloomfield}(1996)}]{Rouzina1996MacroionCloud}%
  \BibitemOpen
  \bibfield  {author} {\bibinfo {author} {\bibfnamefont {I.}~\bibnamefont {Rouzina}}\ and\ \bibinfo {author} {\bibfnamefont {V.~A.}\ \bibnamefont {Bloomfield}},\ }\bibfield  {title} {\enquote {\bibinfo {title} {{Macroion attraction due to electrostatic correlation between screening counterions. 1. Mobile surface-adsorbed ions and diffuse ion cloud}},}\ }\href {\doibase 10.1021/jp960458g} {\bibfield  {journal} {\bibinfo  {journal} {J. Phys. Chem.}\ }\textbf {\bibinfo {volume} {100}},\ \bibinfo {pages} {9977--9989} (\bibinfo {year} {1996})}\BibitemShut {NoStop}%
\bibitem [{\citenamefont {Chu}\ and\ \citenamefont {Wasan}(1996)}]{Chu1996AttractiveParticles}%
  \BibitemOpen
  \bibfield  {author} {\bibinfo {author} {\bibfnamefont {X.}~\bibnamefont {Chu}}\ and\ \bibinfo {author} {\bibfnamefont {D.~T.}\ \bibnamefont {Wasan}},\ }\bibfield  {title} {\enquote {\bibinfo {title} {{Attractive interaction between similarly charged colloidal particles}},}\ }\href {\doibase 10.1006/JCIS.1996.0620} {\bibfield  {journal} {\bibinfo  {journal} {J. Colloid Interface Sci}\ }\textbf {\bibinfo {volume} {184}},\ \bibinfo {pages} {268--278} (\bibinfo {year} {1996})}\BibitemShut {NoStop}%
\bibitem [{\citenamefont {Ha}\ and\ \citenamefont {Liu}(1997)}]{Ha1997Counterion-MediatedRods}%
  \BibitemOpen
  \bibfield  {author} {\bibinfo {author} {\bibfnamefont {B.-Y.}\ \bibnamefont {Ha}}\ and\ \bibinfo {author} {\bibfnamefont {A.~J.}\ \bibnamefont {Liu}},\ }\bibfield  {title} {\enquote {\bibinfo {title} {Counterion-mediated attraction between two like-charged rods},}\ }\href {\doibase 10.1103/PhysRevLett.79.1289} {\bibfield  {journal} {\bibinfo  {journal} {Phys. Rev. Lett.}\ }\textbf {\bibinfo {volume} {79}},\ \bibinfo {pages} {1289--1292} (\bibinfo {year} {1997})}\BibitemShut {NoStop}%
\bibitem [{\citenamefont {Arenzon}, \citenamefont {Stilck},\ and\ \citenamefont {Levin}(1999)}]{Arenzon1999SimplePolyions}%
  \BibitemOpen
  \bibfield  {author} {\bibinfo {author} {\bibfnamefont {J.~J.}\ \bibnamefont {Arenzon}}, \bibinfo {author} {\bibfnamefont {J.~F.}\ \bibnamefont {Stilck}}, \ and\ \bibinfo {author} {\bibfnamefont {Y.}~\bibnamefont {Levin}},\ }\bibfield  {title} {\enquote {\bibinfo {title} {{Simple model for attraction between like-charged polyions}},}\ }\href {\doibase https://doi.org/10.1007/s100510050980} {\bibfield  {journal} {\bibinfo  {journal} {Eur. Phys. J. B}\ }\textbf {\bibinfo {volume} {12}},\ \bibinfo {pages} {79--82} (\bibinfo {year} {1999})}\BibitemShut {NoStop}%
\bibitem [{\citenamefont {Diehl}\ \emph {et~al.}(1999)\citenamefont {Diehl}, \citenamefont {Tamashiro}, \citenamefont {Barbosa},\ and\ \citenamefont {Levin}}]{Diehl1999Density-functionalPlates}%
  \BibitemOpen
  \bibfield  {author} {\bibinfo {author} {\bibfnamefont {A.}~\bibnamefont {Diehl}}, \bibinfo {author} {\bibfnamefont {M.~N.}\ \bibnamefont {Tamashiro}}, \bibinfo {author} {\bibfnamefont {M.~C.}\ \bibnamefont {Barbosa}}, \ and\ \bibinfo {author} {\bibfnamefont {Y.}~\bibnamefont {Levin}},\ }\bibfield  {title} {\enquote {\bibinfo {title} {{Density-functional theory for attraction between like-charged plates}},}\ }\href {\doibase 10.1016/S0378-4371(99)00374-X} {\bibfield  {journal} {\bibinfo  {journal} {Phys. A: Stat. Mech.}\ }\textbf {\bibinfo {volume} {274}},\ \bibinfo {pages} {433--445} (\bibinfo {year} {1999})}\BibitemShut {NoStop}%
\bibitem [{\citenamefont {Grosberg}, \citenamefont {Nguyen},\ and\ \citenamefont {Shklovskii}(2002)}]{Grosberg2002Colloquium:Systems}%
  \BibitemOpen
  \bibfield  {author} {\bibinfo {author} {\bibfnamefont {A.~Y.}\ \bibnamefont {Grosberg}}, \bibinfo {author} {\bibfnamefont {T.~T.}\ \bibnamefont {Nguyen}}, \ and\ \bibinfo {author} {\bibfnamefont {B.~I.}\ \bibnamefont {Shklovskii}},\ }\bibfield  {title} {\enquote {\bibinfo {title} {Colloquium: The physics of charge inversion in chemical and biological systems},}\ }\href {\doibase 10.1103/RevModPhys.74.329} {\bibfield  {journal} {\bibinfo  {journal} {Rev. Mod. Phys.}\ }\textbf {\bibinfo {volume} {74}},\ \bibinfo {pages} {329--345} (\bibinfo {year} {2002})}\BibitemShut {NoStop}%
\bibitem [{\citenamefont {Kudlay}\ and\ \citenamefont {Olvera de~la Cruz}(2003)}]{Kudlay2003PrecipitationSolutions}%
  \BibitemOpen
  \bibfield  {author} {\bibinfo {author} {\bibfnamefont {A.}~\bibnamefont {Kudlay}}\ and\ \bibinfo {author} {\bibfnamefont {M.}~\bibnamefont {Olvera de~la Cruz}},\ }\bibfield  {title} {\enquote {\bibinfo {title} {{Precipitation of oppositely charged polyelectrolytes in salt solutions}},}\ }\href {\doibase 10.1063/1.1629271} {\bibfield  {journal} {\bibinfo  {journal} {J. Chem. Phys.}\ }\textbf {\bibinfo {volume} {120}},\ \bibinfo {pages} {404--412} (\bibinfo {year} {2003})}\BibitemShut {NoStop}%
\bibitem [{\citenamefont {Netz}\ and\ \citenamefont {Orland}(2000)}]{Netz2000BeyondFunctions}%
  \BibitemOpen
  \bibfield  {author} {\bibinfo {author} {\bibfnamefont {R.~R.}\ \bibnamefont {Netz}}\ and\ \bibinfo {author} {\bibfnamefont {H.}~\bibnamefont {Orland}},\ }\bibfield  {title} {\enquote {\bibinfo {title} {{Beyond Poisson-Boltzmann: Fluctuation effects and correlation functions}},}\ }\href {https://link.springer.com/content/pdf/10.1007%2Fs101890050023.pdf} {\bibfield  {journal} {\bibinfo  {journal} {Eur. Phys. J. E}\ }\textbf {\bibinfo {volume} {1}},\ \bibinfo {pages} {203--214} (\bibinfo {year} {2000})}\BibitemShut {NoStop}%
\bibitem [{\citenamefont {Buyukdagli}(2017)}]{Buyukdagli2017Like-chargeMolecules}%
  \BibitemOpen
  \bibfield  {author} {\bibinfo {author} {\bibfnamefont {S.}~\bibnamefont {Buyukdagli}},\ }\bibfield  {title} {\enquote {\bibinfo {title} {{Like-charge attraction and opposite-charge decomplexation between polymers and DNA molecules}},}\ }\href {\doibase 10.1103/PhysRevE.95.022502} {\bibfield  {journal} {\bibinfo  {journal} {Phys. Rev. E}\ }\textbf {\bibinfo {volume} {95}},\ \bibinfo {pages} {22502} (\bibinfo {year} {2017})}\BibitemShut {NoStop}%
\bibitem [{\citenamefont {Suda}, \citenamefont {Suematsu},\ and\ \citenamefont {Akiyama}(2021)}]{Suematsu2018CiteAs}%
  \BibitemOpen
  \bibfield  {author} {\bibinfo {author} {\bibfnamefont {K.}~\bibnamefont {Suda}}, \bibinfo {author} {\bibfnamefont {A.}~\bibnamefont {Suematsu}}, \ and\ \bibinfo {author} {\bibfnamefont {R.}~\bibnamefont {Akiyama}},\ }\bibfield  {title} {\enquote {\bibinfo {title} {{Lateral depletion effect on two-dimensional ordering of bacteriorhodopsins in a lipid bilayer: A theoretical study based on a binary hard-disk model}},}\ }\href {\doibase 10.1063/5.0044399} {\bibfield  {journal} {\bibinfo  {journal} {J. Chem. Phys.}\ }\textbf {\bibinfo {volume} {154}} (\bibinfo {year} {2021}),\ 10.1063/5.0044399},\ \bibinfo {note} {204904},\ \Eprint {http://arxiv.org/abs/https://pubs.aip.org/aip/jcp/article-pdf/doi/10.1063/5.0044399/14007121/204904\_1\_online.pdf} {https://pubs.aip.org/aip/jcp/article-pdf/doi/10.1063/5.0044399/14007121/204904\_1\_online.pdf} \BibitemShut {NoStop}%
\bibitem [{\citenamefont {Misra}\ \emph {et~al.}(2019)\citenamefont {Misra}, \citenamefont {De~Souza}, \citenamefont {Blankschtein},\ and\ \citenamefont {Bazant}}]{Misra2019TheoryElectrolytes}%
  \BibitemOpen
  \bibfield  {author} {\bibinfo {author} {\bibfnamefont {R.~P.}\ \bibnamefont {Misra}}, \bibinfo {author} {\bibfnamefont {J.~P.}\ \bibnamefont {De~Souza}}, \bibinfo {author} {\bibfnamefont {D.}~\bibnamefont {Blankschtein}}, \ and\ \bibinfo {author} {\bibfnamefont {M.~Z.}\ \bibnamefont {Bazant}},\ }\bibfield  {title} {\enquote {\bibinfo {title} {{Theory of Surface Forces in Multivalent Electrolytes}},}\ }\href {\doibase 10.1021/acs.langmuir.9b01110} {\bibfield  {journal} {\bibinfo  {journal} {Langmuir}\ }\textbf {\bibinfo {volume} {35}},\ \bibinfo {pages} {11550--11565} (\bibinfo {year} {2019})}\BibitemShut {NoStop}%
\bibitem [{\citenamefont {Chen}, \citenamefont {Chen},\ and\ \citenamefont {Yang}(2020)}]{Chen2020MultivalentPolyelectrolytes}%
  \BibitemOpen
  \bibfield  {author} {\bibinfo {author} {\bibfnamefont {X.}~\bibnamefont {Chen}}, \bibinfo {author} {\bibfnamefont {E.-Q.}\ \bibnamefont {Chen}}, \ and\ \bibinfo {author} {\bibfnamefont {S.}~\bibnamefont {Yang}},\ }\bibfield  {title} {\enquote {\bibinfo {title} {Multivalent counterions induced attraction between dna polyelectrolytes},}\ }\href@noop {} {\bibfield  {journal} {\bibinfo  {journal} {RSC Adv.}\ }\textbf {\bibinfo {volume} {10}},\ \bibinfo {pages} {1890--1900} (\bibinfo {year} {2020})}\BibitemShut {NoStop}%
\bibitem [{\citenamefont {Gupta}\ \emph {et~al.}(2020{\natexlab{a}})\citenamefont {Gupta}, \citenamefont {Govind~Rajan}, \citenamefont {Carter},\ and\ \citenamefont {Stone}}]{gupta_prl_overcharging}%
  \BibitemOpen
  \bibfield  {author} {\bibinfo {author} {\bibfnamefont {A.}~\bibnamefont {Gupta}}, \bibinfo {author} {\bibfnamefont {A.}~\bibnamefont {Govind~Rajan}}, \bibinfo {author} {\bibfnamefont {E.~A.}\ \bibnamefont {Carter}}, \ and\ \bibinfo {author} {\bibfnamefont {H.~A.}\ \bibnamefont {Stone}},\ }\bibfield  {title} {\enquote {\bibinfo {title} {Ionic layering and overcharging in electrical double layers in a poisson-boltzmann model},}\ }\href {\doibase 10.1103/PhysRevLett.125.188004} {\bibfield  {journal} {\bibinfo  {journal} {Phys. Rev. Lett.}\ }\textbf {\bibinfo {volume} {125}},\ \bibinfo {pages} {188004} (\bibinfo {year} {2020}{\natexlab{a}})}\BibitemShut {NoStop}%
\bibitem [{\citenamefont {Gupta}\ \emph {et~al.}(2020{\natexlab{b}})\citenamefont {Gupta}, \citenamefont {Govind~Rajan}, \citenamefont {Carter},\ and\ \citenamefont {Stone}}]{gupta_force}%
  \BibitemOpen
  \bibfield  {author} {\bibinfo {author} {\bibfnamefont {A.}~\bibnamefont {Gupta}}, \bibinfo {author} {\bibfnamefont {A.}~\bibnamefont {Govind~Rajan}}, \bibinfo {author} {\bibfnamefont {E.~A.}\ \bibnamefont {Carter}}, \ and\ \bibinfo {author} {\bibfnamefont {H.~A.}\ \bibnamefont {Stone}},\ }\bibfield  {title} {\enquote {\bibinfo {title} {Thermodynamics of electrical double layers with electrostatic correlations},}\ }\href {\doibase 10.1021/acs.jpcc.0c08554} {\bibfield  {journal} {\bibinfo  {journal} {J. Phys. Chem. C}\ }\textbf {\bibinfo {volume} {124}},\ \bibinfo {pages} {26830--26842} (\bibinfo {year} {2020}{\natexlab{b}})}\BibitemShut {NoStop}%
\bibitem [{\citenamefont {Hatlo}\ and\ \citenamefont {Lue}(2009)}]{Hatlo2009ACouplings}%
  \BibitemOpen
  \bibfield  {author} {\bibinfo {author} {\bibfnamefont {M.~M.}\ \bibnamefont {Hatlo}}\ and\ \bibinfo {author} {\bibfnamefont {L.}~\bibnamefont {Lue}},\ }\bibfield  {title} {\enquote {\bibinfo {title} {{A field theory for ions near charged surfaces valid from weak to strong couplings}},}\ }\href {\doibase 10.1039/b815578j} {\bibfield  {journal} {\bibinfo  {journal} {Soft Matter}\ }\textbf {\bibinfo {volume} {5}},\ \bibinfo {pages} {125--133} (\bibinfo {year} {2009})}\BibitemShut {NoStop}%
\bibitem [{\citenamefont {Zhou}(2019)}]{ShiqiZhou2019EffectiveLike}%
  \BibitemOpen
  \bibfield  {author} {\bibinfo {author} {\bibfnamefont {S.}~\bibnamefont {Zhou}},\ }\bibfield  {title} {\enquote {\bibinfo {title} {Effective electrostatic potential between two oppositely charged cylinder rods in primitive model and extended primitive model electrolytes},}\ }\href {\doibase 10.1088/1742-5468/ab00e1} {\bibfield  {journal} {\bibinfo  {journal} {J. Stat. Mech. Theory Exp.}\ }\textbf {\bibinfo {volume} {2019}},\ \bibinfo {pages} {033213} (\bibinfo {year} {2019})}\BibitemShut {NoStop}%
\bibitem [{\citenamefont {Ferrari}\ \emph {et~al.}(2010)\citenamefont {Ferrari}, \citenamefont {Kaufmann}, \citenamefont {Winnefeld},\ and\ \citenamefont {Plank}}]{Ferrari2010InteractionMeasurements}%
  \BibitemOpen
  \bibfield  {author} {\bibinfo {author} {\bibfnamefont {L.}~\bibnamefont {Ferrari}}, \bibinfo {author} {\bibfnamefont {J.}~\bibnamefont {Kaufmann}}, \bibinfo {author} {\bibfnamefont {F.}~\bibnamefont {Winnefeld}}, \ and\ \bibinfo {author} {\bibfnamefont {J.}~\bibnamefont {Plank}},\ }\bibfield  {title} {\enquote {\bibinfo {title} {{Interaction of cement model systems with superplasticizers investigated by atomic force microscopy, zeta potential, and adsorption measurements}},}\ }\href {\doibase 10.1016/J.JCIS.2010.03.005} {\bibfield  {journal} {\bibinfo  {journal} {J. Colloid Interface Sci.}\ }\textbf {\bibinfo {volume} {347}},\ \bibinfo {pages} {15--24} (\bibinfo {year} {2010})}\BibitemShut {NoStop}%
\bibitem [{\citenamefont {Varennes}\ and\ \citenamefont {{Van De Ven}}(1988)}]{Varennes1988EffectsFlow}%
  \BibitemOpen
  \bibfield  {author} {\bibinfo {author} {\bibfnamefont {S.}~\bibnamefont {Varennes}}\ and\ \bibinfo {author} {\bibfnamefont {T.}~\bibnamefont {{Van De Ven}}},\ }\bibfield  {title} {\enquote {\bibinfo {title} {Effects of polyelectrolyte on the deposition and detachment of colloidal particles subjected to flow},}\ }\href {\doibase https://doi.org/10.1016/0166-6622(88)80049-0} {\bibfield  {journal} {\bibinfo  {journal} {Colloids Surf.}\ }\textbf {\bibinfo {volume} {33}},\ \bibinfo {pages} {63--74} (\bibinfo {year} {1988})}\BibitemShut {NoStop}%
\bibitem [{\citenamefont {Guzey}\ and\ \citenamefont {McClements}(2006)}]{Guzey2006FormationIndustry}%
  \BibitemOpen
  \bibfield  {author} {\bibinfo {author} {\bibfnamefont {D.}~\bibnamefont {Guzey}}\ and\ \bibinfo {author} {\bibfnamefont {D.~J.}\ \bibnamefont {McClements}},\ }\bibfield  {title} {\enquote {\bibinfo {title} {{Formation, stability and properties of multilayer emulsions for application in the food industry}},}\ }\href {\doibase 10.1016/J.CIS.2006.11.021} {\bibfield  {journal} {\bibinfo  {journal} {Adv. Colloid Interface Sci.}\ }\textbf {\bibinfo {volume} {128-130}},\ \bibinfo {pages} {227--248} (\bibinfo {year} {2006})}\BibitemShut {NoStop}%
\bibitem [{\citenamefont {Agheli}\ \emph {et~al.}(2006)\citenamefont {Agheli}, \citenamefont {Malmstr{\"{o}}m}, \citenamefont {Larsson}, \citenamefont {Textor},\ and\ \citenamefont {Sutherland}}]{Agheli2006LargeApplications}%
  \BibitemOpen
  \bibfield  {author} {\bibinfo {author} {\bibfnamefont {H.}~\bibnamefont {Agheli}}, \bibinfo {author} {\bibfnamefont {J.}~\bibnamefont {Malmstr{\"{o}}m}}, \bibinfo {author} {\bibfnamefont {E.~M.}\ \bibnamefont {Larsson}}, \bibinfo {author} {\bibfnamefont {M.}~\bibnamefont {Textor}}, \ and\ \bibinfo {author} {\bibfnamefont {D.~S.}\ \bibnamefont {Sutherland}},\ }\bibfield  {title} {\enquote {\bibinfo {title} {{Large area protein nanopatterning for biological applications}},}\ }\href {\doibase 10.1021/NL060403I/ASSET/IMAGES/LARGE/NL060403IF00005.JPEG} {\bibfield  {journal} {\bibinfo  {journal} {Nano Letters}\ }\textbf {\bibinfo {volume} {6}},\ \bibinfo {pages} {1165--1171} (\bibinfo {year} {2006})}\BibitemShut {NoStop}%
\bibitem [{\citenamefont {Aranda-Espinoza}\ \emph {et~al.}(1999)\citenamefont {Aranda-Espinoza}, \citenamefont {Chen}, \citenamefont {Dan}, \citenamefont {Lubensky}, \citenamefont {Nelson}, \citenamefont {Ramos},\ and\ \citenamefont {Weitz}}]{Aranda-Espinoza1999ElectrostaticObjects}%
  \BibitemOpen
  \bibfield  {author} {\bibinfo {author} {\bibfnamefont {H.}~\bibnamefont {Aranda-Espinoza}}, \bibinfo {author} {\bibfnamefont {Y.}~\bibnamefont {Chen}}, \bibinfo {author} {\bibfnamefont {N.}~\bibnamefont {Dan}}, \bibinfo {author} {\bibfnamefont {T.~C.}\ \bibnamefont {Lubensky}}, \bibinfo {author} {\bibfnamefont {P.}~\bibnamefont {Nelson}}, \bibinfo {author} {\bibfnamefont {L.}~\bibnamefont {Ramos}}, \ and\ \bibinfo {author} {\bibfnamefont {D.~A.}\ \bibnamefont {Weitz}},\ }\bibfield  {title} {\enquote {\bibinfo {title} {{Electrostatic repulsion of positively charged vesicles and negatively charged objects}},}\ }\href {\doibase 10.1126/SCIENCE.285.5426.394/ASSET/5C55B651-74BC-44D5-9605-C1467D49FC4E/ASSETS/GRAPHIC/SE2797679003.JPEG} {\bibfield  {journal} {\bibinfo  {journal} {Science}\ }\textbf {\bibinfo {volume} {285}},\ \bibinfo {pages} {394--397} (\bibinfo {year} {1999})}\BibitemShut {NoStop}%
\bibitem [{\citenamefont {Schiessel}(2003)}]{chromatinphysics}%
  \BibitemOpen
  \bibfield  {author} {\bibinfo {author} {\bibfnamefont {H.}~\bibnamefont {Schiessel}},\ }\bibfield  {title} {\enquote {\bibinfo {title} {The physics of chromatin},}\ }\href {\doibase 10.1088/0953-8984/15/19/203} {\bibfield  {journal} {\bibinfo  {journal} {J. Phys. Condens. Matter}\ }\textbf {\bibinfo {volume} {15}},\ \bibinfo {pages} {R699} (\bibinfo {year} {2003})}\BibitemShut {NoStop}%
\bibitem [{\citenamefont {Pieters}\ \emph {et~al.}(2015)\citenamefont {Pieters}, \citenamefont {Van~Eldijk}, \citenamefont {Nolte},\ and\ \citenamefont {Mecinovi{\'{c}}}}]{Pieters2015NaturalAssemblies}%
  \BibitemOpen
  \bibfield  {author} {\bibinfo {author} {\bibfnamefont {B.~J.}\ \bibnamefont {Pieters}}, \bibinfo {author} {\bibfnamefont {M.~B.}\ \bibnamefont {Van~Eldijk}}, \bibinfo {author} {\bibfnamefont {R.~J.}\ \bibnamefont {Nolte}}, \ and\ \bibinfo {author} {\bibfnamefont {J.}~\bibnamefont {Mecinovi{\'{c}}}},\ }\bibfield  {title} {\enquote {\bibinfo {title} {{Natural supramolecular protein assemblies}},}\ }\href {\doibase 10.1039/C5CS00157A} {\bibfield  {journal} {\bibinfo  {journal} {Chem. Soc. Rev.}\ }\textbf {\bibinfo {volume} {45}},\ \bibinfo {pages} {24--39} (\bibinfo {year} {2015})}\BibitemShut {NoStop}%
\bibitem [{\citenamefont {Gelbart}\ \emph {et~al.}(2007)\citenamefont {Gelbart}, \citenamefont {Bruinsma}, \citenamefont {Pincus},\ and\ \citenamefont {Adrian~Parsegian}}]{Gelbart2007DNAInspiredElectrostatics}%
  \BibitemOpen
  \bibfield  {author} {\bibinfo {author} {\bibfnamefont {W.~M.}\ \bibnamefont {Gelbart}}, \bibinfo {author} {\bibfnamefont {R.~F.}\ \bibnamefont {Bruinsma}}, \bibinfo {author} {\bibfnamefont {P.~A.}\ \bibnamefont {Pincus}}, \ and\ \bibinfo {author} {\bibfnamefont {V.}~\bibnamefont {Adrian~Parsegian}},\ }\bibfield  {title} {\enquote {\bibinfo {title} {{DNA‐Inspired Electrostatics}},}\ }\href {\doibase 10.1063/1.1325230} {\bibfield  {journal} {\bibinfo  {journal} {Phys. Today}\ }\textbf {\bibinfo {volume} {53}},\ \bibinfo {pages} {38} (\bibinfo {year} {2007})}\BibitemShut {NoStop}%
\bibitem [{\citenamefont {Solis}\ and\ \citenamefont {Cruz}(2007)}]{Solis2007FlexibleCounterattract}%
  \BibitemOpen
  \bibfield  {author} {\bibinfo {author} {\bibfnamefont {F.~J.}\ \bibnamefont {Solis}}\ and\ \bibinfo {author} {\bibfnamefont {M.~O. d.~l.}\ \bibnamefont {Cruz}},\ }\bibfield  {title} {\enquote {\bibinfo {title} {{Flexible Polymers Also Counterattract}},}\ }\href {\doibase 10.1063/1.1349627} {\bibfield  {journal} {\bibinfo  {journal} {Phys. Today}\ }\textbf {\bibinfo {volume} {54}},\ \bibinfo {pages} {71} (\bibinfo {year} {2007})}\BibitemShut {NoStop}%
\bibitem [{\citenamefont {Agrawal}\ \emph {et~al.}(2023)\citenamefont {Agrawal}, \citenamefont {Kaur}, \citenamefont {Carraro},\ and\ \citenamefont {Wang}}]{agrawal2023lca}%
  \BibitemOpen
  \bibfield  {author} {\bibinfo {author} {\bibfnamefont {N.~R.}\ \bibnamefont {Agrawal}}, \bibinfo {author} {\bibfnamefont {R.}~\bibnamefont {Kaur}}, \bibinfo {author} {\bibfnamefont {C.}~\bibnamefont {Carraro}}, \ and\ \bibinfo {author} {\bibfnamefont {R.}~\bibnamefont {Wang}},\ }\bibfield  {title} {\enquote {\bibinfo {title} {{Ion correlation-driven like-charge attraction in multivalent salt solutions}},}\ }\href {\doibase 10.1063/5.0181061} {\bibfield  {journal} {\bibinfo  {journal} {J. Chem. Phys.}\ }\textbf {\bibinfo {volume} {159}},\ \bibinfo {pages} {244905} (\bibinfo {year} {2023})},\ \Eprint {http://arxiv.org/abs/https://pubs.aip.org/aip/jcp/article-pdf/doi/10.1063/5.0181061/18277753/244905\_1\_5.0181061.pdf} {https://pubs.aip.org/aip/jcp/article-pdf/doi/10.1063/5.0181061/18277753/244905\_1\_5.0181061.pdf} \BibitemShut {NoStop}%
\bibitem [{\citenamefont {Agrawal}, \citenamefont {Duan},\ and\ \citenamefont {Wang}(2024)}]{Agrawal2022OnLayers}%
  \BibitemOpen
  \bibfield  {author} {\bibinfo {author} {\bibfnamefont {N.~R.}\ \bibnamefont {Agrawal}}, \bibinfo {author} {\bibfnamefont {C.}~\bibnamefont {Duan}}, \ and\ \bibinfo {author} {\bibfnamefont {R.}~\bibnamefont {Wang}},\ }\bibfield  {title} {\enquote {\bibinfo {title} {Nature of overcharging and charge inversion in electrical double layers},}\ }\href {\doibase 10.1021/acs.jpcb.3c04739} {\bibfield  {journal} {\bibinfo  {journal} {J. Phys. Chem. B}\ }\textbf {\bibinfo {volume} {128}},\ \bibinfo {pages} {303--311} (\bibinfo {year} {2024})},\ \Eprint {http://arxiv.org/abs/https://doi.org/10.1021/acs.jpcb.3c04739} {https://doi.org/10.1021/acs.jpcb.3c04739} \BibitemShut {NoStop}%
\bibitem [{\citenamefont {Agrawal}\ and\ \citenamefont {Wang}(2023)}]{agrawal_aiche}%
  \BibitemOpen
  \bibfield  {author} {\bibinfo {author} {\bibfnamefont {N.~R.}\ \bibnamefont {Agrawal}}\ and\ \bibinfo {author} {\bibfnamefont {R.}~\bibnamefont {Wang}},\ }\bibfield  {title} {\enquote {\bibinfo {title} {Non-monotonic salt concentration dependence of inverted electrokinetic flow},}\ }\href {\doibase https://doi.org/10.1002/aic.18269} {\bibfield  {journal} {\bibinfo  {journal} {AIChE Journal}\ ,\ \bibinfo {pages} {e18269}} (\bibinfo {year} {2023})}\BibitemShut {NoStop}%
\bibitem [{\citenamefont {Van Der~Heyden}\ \emph {et~al.}(2006)\citenamefont {Van Der~Heyden}, \citenamefont {Stein}, \citenamefont {Besteman}, \citenamefont {Lemay},\ and\ \citenamefont {Dekker}}]{VanDerHeyden2006ChargeCurrents}%
  \BibitemOpen
  \bibfield  {author} {\bibinfo {author} {\bibfnamefont {F.~H.}\ \bibnamefont {Van Der~Heyden}}, \bibinfo {author} {\bibfnamefont {D.}~\bibnamefont {Stein}}, \bibinfo {author} {\bibfnamefont {K.}~\bibnamefont {Besteman}}, \bibinfo {author} {\bibfnamefont {S.~G.}\ \bibnamefont {Lemay}}, \ and\ \bibinfo {author} {\bibfnamefont {C.}~\bibnamefont {Dekker}},\ }\bibfield  {title} {\enquote {\bibinfo {title} {{Charge inversion at high ionic strength studied by streaming currents}},}\ }\href {\doibase 10.1103/PhysRevLett.96.224502} {\bibfield  {journal} {\bibinfo  {journal} {Phys. Rev. Lett.}\ }\textbf {\bibinfo {volume} {96}},\ \bibinfo {pages} {224502} (\bibinfo {year} {2006})}\BibitemShut {NoStop}%
\bibitem [{\citenamefont {Hsiao}\ and\ \citenamefont {Luijten}(2006)}]{Hsiao2006Salt-inducedPolyelectrolytes}%
  \BibitemOpen
  \bibfield  {author} {\bibinfo {author} {\bibfnamefont {P.~Y.}\ \bibnamefont {Hsiao}}\ and\ \bibinfo {author} {\bibfnamefont {E.}~\bibnamefont {Luijten}},\ }\bibfield  {title} {\enquote {\bibinfo {title} {{Salt-induced collapse and reexpansion of highly charged flexible polyelectrolytes}},}\ }\href {\doibase 10.1103/PhysRevLett.97.148301} {\bibfield  {journal} {\bibinfo  {journal} {Phys. Rev. Lett.}\ }\textbf {\bibinfo {volume} {97}},\ \bibinfo {pages} {148301} (\bibinfo {year} {2006})}\BibitemShut {NoStop}%
\bibitem [{\citenamefont {Levin}(2002)}]{Levin2002ElectrostaticBiology}%
  \BibitemOpen
  \bibfield  {author} {\bibinfo {author} {\bibfnamefont {Y.}~\bibnamefont {Levin}},\ }\bibfield  {title} {\enquote {\bibinfo {title} {{Electrostatic correlations: From plasma to biology}},}\ }\href {\doibase 10.1088/0034-4885/65/11/201} {\bibfield  {journal} {\bibinfo  {journal} {Rep. Prog. Phys.}\ }\textbf {\bibinfo {volume} {65}},\ \bibinfo {pages} {1577--1632} (\bibinfo {year} {2002})}\BibitemShut {NoStop}%
\bibitem [{\citenamefont {Agrawal}\ and\ \citenamefont {Wang}(2022)}]{Agrawal2022Self-ConsistentFluids}%
  \BibitemOpen
  \bibfield  {author} {\bibinfo {author} {\bibfnamefont {N.~R.}\ \bibnamefont {Agrawal}}\ and\ \bibinfo {author} {\bibfnamefont {R.}~\bibnamefont {Wang}},\ }\bibfield  {title} {\enquote {\bibinfo {title} {Self-consistent description of vapor-liquid interface in ionic fluids},}\ }\href {\doibase 10.1103/PhysRevLett.129.228001} {\bibfield  {journal} {\bibinfo  {journal} {Phys. Rev. Lett.}\ }\textbf {\bibinfo {volume} {129}},\ \bibinfo {pages} {228001} (\bibinfo {year} {2022})}\BibitemShut {NoStop}%
\bibitem [{\citenamefont {Netz}\ and\ \citenamefont {Orland}(2003)}]{Netz2003VariationalSystems}%
  \BibitemOpen
  \bibfield  {author} {\bibinfo {author} {\bibfnamefont {R.~R.}\ \bibnamefont {Netz}}\ and\ \bibinfo {author} {\bibfnamefont {H.}~\bibnamefont {Orland}},\ }\bibfield  {title} {\enquote {\bibinfo {title} {{Variational charge renormalization in charged systems}},}\ }\href {\doibase 10.1140/epje/i2002-10159-0} {\bibfield  {journal} {\bibinfo  {journal} {Eur. Phys. J. E}\ }\textbf {\bibinfo {volume} {11}},\ \bibinfo {pages} {301--311} (\bibinfo {year} {2003})}\BibitemShut {NoStop}%
\bibitem [{\citenamefont {Wang}(2010)}]{Wang2010FluctuationEnergy}%
  \BibitemOpen
  \bibfield  {author} {\bibinfo {author} {\bibfnamefont {Z.~G.}\ \bibnamefont {Wang}},\ }\bibfield  {title} {\enquote {\bibinfo {title} {{Fluctuation in electrolyte solutions: The self energy}},}\ }\href {\doibase 10.1103/PhysRevE.81.021501} {\bibfield  {journal} {\bibinfo  {journal} {Phys. Rev. E}\ }\textbf {\bibinfo {volume} {81}},\ \bibinfo {pages} {021501} (\bibinfo {year} {2010})}\BibitemShut {NoStop}%
\bibitem [{\citenamefont {Wang}\ and\ \citenamefont {Wang}(2015)}]{Wang2015OnSurfaces}%
  \BibitemOpen
  \bibfield  {author} {\bibinfo {author} {\bibfnamefont {R.}~\bibnamefont {Wang}}\ and\ \bibinfo {author} {\bibfnamefont {Z.~G.}\ \bibnamefont {Wang}},\ }\bibfield  {title} {\enquote {\bibinfo {title} {{On the theoretical description of weakly charged surfaces}},}\ }\href {\doibase 10.1063/1.4914170} {\bibfield  {journal} {\bibinfo  {journal} {J. Chem. Phys.}\ }\textbf {\bibinfo {volume} {142}},\ \bibinfo {pages} {104705} (\bibinfo {year} {2015})}\BibitemShut {NoStop}%
\bibitem [{\citenamefont {Xu}\ and\ \citenamefont {Maggs}(2014)}]{Xu2014SolvingEquations}%
  \BibitemOpen
  \bibfield  {author} {\bibinfo {author} {\bibfnamefont {Z.}~\bibnamefont {Xu}}\ and\ \bibinfo {author} {\bibfnamefont {A.~C.}\ \bibnamefont {Maggs}},\ }\bibfield  {title} {\enquote {\bibinfo {title} {{Solving fluctuation-enhanced Poisson-Boltzmann equations}},}\ }\href {\doibase 10.1016/j.jcp.2014.07.004} {\bibfield  {journal} {\bibinfo  {journal} {J. Comput. Phys.}\ }\textbf {\bibinfo {volume} {275}},\ \bibinfo {pages} {310--322} (\bibinfo {year} {2014})}\BibitemShut {NoStop}%
\bibitem [{\citenamefont {{Burns}}\ \emph {et~al.}(2020)\citenamefont {{Burns}}, \citenamefont {{Vasil}}, \citenamefont {{Oishi}}, \citenamefont {{Lecoanet}},\ and\ \citenamefont {{Brown}}}]{burns_dedalus}%
  \BibitemOpen
  \bibfield  {author} {\bibinfo {author} {\bibfnamefont {K.~J.}\ \bibnamefont {{Burns}}}, \bibinfo {author} {\bibfnamefont {G.~M.}\ \bibnamefont {{Vasil}}}, \bibinfo {author} {\bibfnamefont {J.~S.}\ \bibnamefont {{Oishi}}}, \bibinfo {author} {\bibfnamefont {D.}~\bibnamefont {{Lecoanet}}}, \ and\ \bibinfo {author} {\bibfnamefont {B.~P.}\ \bibnamefont {{Brown}}},\ }\bibfield  {title} {\enquote {\bibinfo {title} {{Dedalus: A flexible framework for numerical simulations with spectral methods}},}\ }\href {\doibase 10.1103/PhysRevResearch.2.023068} {\bibfield  {journal} {\bibinfo  {journal} {Phys. Rev. Res.}\ }\textbf {\bibinfo {volume} {2}},\ \bibinfo {eid} {023068} (\bibinfo {year} {2020})},\ \Eprint {http://arxiv.org/abs/1905.10388} {arXiv:1905.10388 [astro-ph.IM]} \BibitemShut {NoStop}%
\end{thebibliography}%

\end{document}